# Machine Learning Aided Static Malware Analysis: A Survey and Tutorial

Andrii Shalaginov, Sergii Banin, Ali Dehghantanha, Katrin Franke

**Abstract** Malware analysis and detection techniques have been evolving during the last decade as a reflection to development of different malware techniques to evade network-based and host-based security protections. The fast growth in variety and number of malware species made it very difficult for forensics investigators to provide an on time response. Therefore, Machine Learning (ML) aided malware analysis became a necessity to automate different aspects of static and dynamic malware investigation. We believe that machine learning aided static analysis can be used as a methodological approach in technical Cyber Threats Intelligence (CTI) rather than resource-consuming dynamic malware analysis that has been thoroughly studied before. In this paper, we address this research gap by conducting an in-depth survey of different machine learning methods for classification of static characteristics of 32-bit malicious Portable Executable (PE32) Windows files and develop taxonomy for better understanding of these techniques. Afterwards, we offer a tutorial on how different machine learning techniques can be utilized in extraction and analysis of a variety of static characteristic of PE binaries and evaluate accuracy and practical generalization of these techniques. Finally, the results of experimental

Andrii Shalaginov
Norwegian Information Security Laboratory, Center for Cyber- and Information Security, Norwegian University of Science and Technology, Gjvik, Norway, e-mail: andrii.shalaginov@ntnu.no

Sergii Banin
Norwegian Information Security Laboratory, Center for Cyber- and Information Security, Norwegian University of Science and Technology, Gjvik, Norway,e-mail: sergii.banin@ntnu.no

Ali Dehghantanha
School of Computing, Science and Engineering, University of Salford, UK e-mail: a.dehghantanha@salford.ac.uk

Katrin Franke
Norwegian Information Security Laboratory, Center for Cyber- and Information Security, Norwegian University of Science and Technology, Gjvik, Norway, e-mail: katrin.franke@ntnu.no

study of all the method using common data was given to demonstrate the accuracy and complexity. This paper may serve as a stepping stone for future researchers in cross-disciplinary field of machine learning aided malware forensics.

## 1 Introduction

Stealing users' personal and private information has been always among top interests of malicious programs [8]. Platforms which are widely used by normal users have always been best targets for malware developers [9].

Attackers have leveraged malware to target personal computers [22], mobile devices [61], cloud storage systems [13], Supervisory Control and Data Acquisition Systems (SCADA) [12], Internet of Things (IoT) network [81] and even big data platforms [67].

Forensics examiners and incident handlers on the other side have developed different techniques for detection of compromised systems, removal of detected malicious programs [23, 14], network traffic [63], and even log analysis [69]. Different models have been suggested for detection, correlation and analyses of cyber threats [20] (on a range of mobile devices [43] and mobile applications [45], cloud applications [11], cloud infrastructure [46] and Internet of Things networks [47]). Windows users are still comprising majority of Internet users hence, it is not surprising to see Windows as the most adopted PC Operating System (OS) on top of the list of malware targeted platforms [73]. In response, lots of efforts have been made to secure Windows platform such as educating users [54, 25], embedding an antivirus software [40], deploying anti-malware and anti-exploitation tools [53, 52], and limiting users applications privilege [41].

In spite of all security enhancements, many malware are still successfully compromising Windows machines [36, 73] and malware is still ranked as an important threat to Windows platforms [33]. As result, many security professionals are still required to spend a lot of time on analyzing different malware species [10]. This is a logical step since malware analysis plays a crucial role in Cyber Threats Intelligence (CTI). There has been proposed a portal to facilitate CTI and malware analysis through interactive collaboration and information fusion [56].

There are two major approaches for malware analysis namely *static* (code) and *dynamic* (behavioral) malware analysis [16, 8]. In dynamic malware analysis, samples are executed and their run time behavior such as transmitted network traffic, the length of execution, changes that are made in the file system, etc. are used to understand the malware behavior and create indications of compromise for malware detection [16]. However, dynamic analysis techniques can be easily evaded by malware that are aware of execution conditions and computing environment [34]. Dynamic malware analysis techniques can only provide a snapshot view of the malware behavior and hence very limited in analysis of Polymorph or Metamorph species [44]. Moreover, dynamic malware analysis techniques are quite resource hungry which limits their enterprise deployment [37].

In static malware analysis, the analyst is reversing the malware code to achieve a deeper understanding of the malware possible activities. [28]. Static analysis relies on extraction of a variety of characteristics from the binary file such that function calls, header sections, etc. [83]. Such characteristics may reveal indicators of malicious activity that are going to be used in CTI [57]. However, static analysis is quite a slow process and requires a lot of human interpretation and hence [8].

Static analysis of PE32 is a many-sided challenge that was studied by different authors. Static malware analysis also was used before for discovering interconnections in malware species for improved Cyber Threat Intellifence [42, 66]. As 32-bit malware are still capable of infecting 64-bit platforms and considering there are still many 32-bit Windows OS it is not surprising that still majority of Windows malware are 32-bit Portable Executable files [8]. To authors knowledge there has not been a comparative study of ML-based static malware using a single dataset which produces comparable results. We believe that utilization of ML-aided automated analysis can speed up intelligent malware analysis process and reduce human interaction required for binaries processing. Therefore, there is a need for thorough review of the relevant scientific contributions and offer a taxonomy for automated static malware analysis.

The remainder of this paper is organized as follows. We first offer a comprehensive review of existing literature in machine learning aided static malware analysis. We believe this survey paves the way for further research in application of machine learning in static malware analysis and calls for further development in this field. Then, taxonomy of feature construction methods for variety of static characteristics and corresponding ML classification methods is offered. Afterwards, we offer a tutorial that applies variety of set of machine learning techniques and compares their performance. The tutorial findings provide a clear picture of pros and cons of ML-aided static malware analysis techniques. To equally compare all the methods we used one benign and two malware datasets to evaluate all of the studied methods. This important part complements the paper due to the fact that most of the surveyed works used own collections, sometimes not available for public access or not published at all. Therefore, experimental study showed performance comparison and other practical aspects of ML-aided malware analysis. Section 4 gives an insight into a practical routine that we used to establish our experimental setup. Analysis of results and findings are given in the Section 5. Finally, the paper is concluded and several future works are suggested in the Section 6.

## 2 An overview of Machine Learning-aided static malware detection

This section provides an analysis of detectable static properties of 32 bit PE malware followed by detailed description of different machine learning techniques to develop a taxonomy of machine learning techniques for static malware analysis.

### 2.1 Static characteristics of PE files

PE file format was introduced in Windows 3.1 as PE32 and further developed as PE32+ format for 64 bit Windows Operating Systems. PE files contain a Common Object File Format (COFF) header, standard COFF fields such as header, section table, data directories and Import Address Table (IAT). Beside the PE header fields a number of other static features can be extracted from a binary executable such as strings, entropy and size of various sections.

To be able to apply Machine Learning PE32 files static characteristics should be converted into a machine-understandable features. There exist different types of features depending on the nature of their values such that *numerical* that describes a quantitative measure (can be integer, real or binary value) or *nominal* that describes finite set of categories or labels. An example of the *numerical* feature is CPU (in %) or RAM (in Megabytes) usage, while *nominal* can be a file type (like $*.dll$ or $*.exe$) or Application Program Interface (API) function call (like $write()$ or $read()$).

1. *n-grams of byte sequences* is a well-known method of feature construction utilizing sequences of bytes from binary files to create features. Many tools have been developed for this purpose such as *hexdump* [39] created 4-grams from byte sequences of PE32 files. The features are collected by sliding window of *n* bytes. This resulted in 200 millions of features using 10-grams for about two thousands files in overall. Moreover, feature selection (FS) was applied to select 500 most valuable features based on Information Gain metric. Achieved accuracy on malware detection was up to 97% using such features. Another work on byte n-grams [51] described usage of 100-500 selected n-grams yet on a set of 250 malicious and 250 benign samples. Similar approach [31] was used with $10,\dots,10,000$ best n-grams for $n = 1,\dots,10$. Additionally, ML methods such that Naive Bayes, C4.5, k-NN and others were investigated to evaluate their applicability and accuracy. Finally, a range of 1-8 n-grams [27] can result in 500 best selected n-grams that are used later to train AdaBoost and Random Forests in addition to previously mentioned works.
2. *Opcode sequences* or operation codes are set of consecutive low level machine abstractions used to perform various CPU operations. As it was shown [62] such features can be used to train Machine Learning methods for successful classification of the malware samples. However, there should be a balance between the size of the feature set and the length of n-gram opcode sequence. N-grams with the size of 4 and 5 result in highest classification accuracy as unknown malware samples could be unveiled on a collection of 17,000 malware and 1,000 benign files with a classification accuracy up to 94% [58]. Bragen [6] explored reliability of malware analysis using sequences of opcodes based on the 992 PE-files malware and benign samples. During the experiments, about 50 millions of opcodes were extracted. 1-gram- and 2-gram-based features showed good computational results and accuracy. Wang et al. [79] presented how the 2-tuple opcode sequences can be used in combination with density clustering to detect malicious or benign files.

3. *API calls* are the function calls used by a program to execute specific functionality. We have to distinguish between System API calls that are available through standard system DLLs and User API calls provided by user installed software. These are designed to perform a pre-defined task during invocation. Suspicious API calls, anti-VM and anti-debugger hooks and calls can be extracted by PE analysers such as PEframe [4]. [83] studied 23 malware samples and found that some of the API calls are present only in malwares rather than benign software. Function calls may compose in graphs to represent PE32 header features as nodes, edges and subgraphs [84]. This work shows that ML methods achieve accuracy of 96% on 24 features extracted after analysis of 1,037 malware and 2,072 benign executables. Further, in [71] 20,682 API calls were extracted using PE parser for 1,593 malicious and benign samples. Such large number of extracted features can help to create linearly separable model that is crucial for many ML methods as Support Vector Machines (SVM) or single-layer Neural Networks. Another work by [55] described how API sequences can be analysed in analogy with byte n-grams and opcode n-grams to extract corresponding features to classify malware and benign files. Also in this work, an array of API calls from IAT (PE32 header filed) was processed by Fisher score to select relevant features after analysis of more than 34k samples.
4. *PE header* represents a collection of meta data related to a Portable Executable file. Basic features that can be extracted from PE32 header are *Size of Header*, *Size of Uninitialized Data*, *Size of Stack Reserve*, which may indicate if a binary file is malicious or benign [15]. The work [76] utilized Decision Trees to analyse PE header structural information for describing malicious and benign files. [77] used 125 raw header characteristics, 31 section characteristics, 29 section characteristics to detect unknown malware in a semi-supervised approach. Another work [80] used a dataset containing 7,863 malware samples from Vx Heaven web site in addition to 1,908 benign files to develop a SVM based malware detection model with accuracy of 98%. [38] used F-score as a performance metric to analyse PE32 header features of 164,802 malicious and benign samples. Also [29] presented research of two novel methods related to PE32 string-based classier that do not require additional extraction of structural or meta-data information from the binary files. Moreover, [84] described application of 24 features along with API calls for classification of malware and benign samples from Vx-Heaven and Windows XP SP3 respectively. Further, ensemble of features was explored [59], where authors used in total 209 features including structural and raw data from PE32 file header. Further, Le-Khac et al. [35] focused on Control Flow Change over first 256 addresses to construct n-gram features.

In addition to study of specific features used for malware detection, we analyzed articles devoted to application of ML for static malware analysis published between 2000 and 2016, which covers the timeline of Windows NT family that are still in use as depicted in the Figure 1. We can see that the number of papers that are relevant to our study is growing from 2009 and later, which can be justified on the basis of increase in the number of Windows users (potential targets) and corresponding malware families.

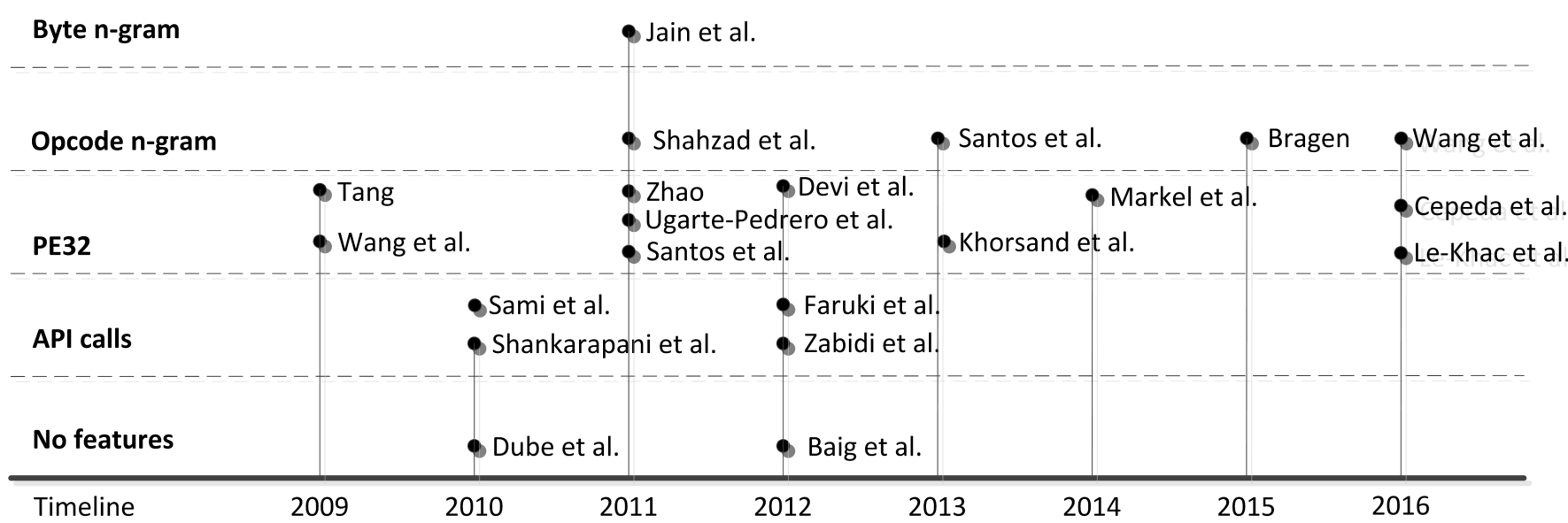

Fig. 1: Timeline of works since 2009 that involved static analysis of Portable Executable files using method characteristics using also ML method for binary malware classification

**Challenges.** Despite the fact that some of the feature construction techniques reflected promising precision of 90+ % in differentiation between malicious and benign executables, there are still no best static characteristic that guarantee 100% accuracy of malware detection. This can be explained by the fact that malware are using obfuscation and encryption techniques to subvert detection mechanisms. In addition, more accurate approaches such as bytes N-GRAMS are quite resource intensive and hardly practical in the real world.

## 2.2 *Machine Learning methods used for static-based malware detection*

### 2.2.1 Statistical methods

Exploring large amounts of binary files consists of statistical features may be simplified using so called frequencies or likelihood of features values. These methods are made to provide prediction about the binary executable class based on statistics of different static characteristics (either automatically or manually collected) which are applicable to malware analysis too as describe by Shabtai et al. [60]. To process such data, extract new or make predictions the following set of statistical methods can be used:

**Naive Bayes** is a simple probabilistic classifier, which is based on Bayesian theorem with naive assumptions about independence in correlation of different features. The Bayes Rule can be explained as a following conditional independences of features values with respect to a class:

$$P(C_k|V) = P(C_k)\frac{P(V|C_k)}{P(V)} \tag{1}$$

where $P(C_k)$ is a prior probability of class $C_k, k = 1, ..., m_0$ which is calculated from collected statistics according to description of variables provided by Kononenko et al. [32]. This method is considered to tackle just binary classification problem (benign against malicious) since it was originally designed as multinomial classifier. $V = \langle v_1, ..., v_a \rangle$ is the vector of attributes values that belongs to a sample. In case of Naive Bayes *input* values should be symbolical, for example strings, opcodes, instruction n-grams etc. $P(V)$ is the prior probability of a sample described with vector $V$. Having training data set and given vector $V$ we count how many samples contain equal values of attributes (e.g. based on the number of sections or given opcode sequence). It is important to mention that $V$ have not to be of length of full attribute vector and can contain only one attribute value. $P(V|C_k)$ is the conditional probability of a sample described with $V$ given the class $C_k$. And $P(C_k|V)$ conditional probability of class $C_k$ with $V$. Based on simple probability theory we can describe conditional independence of attribute values $v_i$ given the class $C_k$:

$$P(V|C_k) = P(v_1 \wedge ... \wedge v_a|C_k) = \prod_{i=1}^{a} P(v_i|C_k) \tag{2}$$

Dropping the mathematical operations we get final version of Equation 1:

$$P(C_k|V) = P(C_k) \prod_{i=1}^{a} \frac{P(C_k|v_i)}{P(C_k)} \tag{3}$$

So, the task of this machine learning algorithm is to calculate conditional and unconditional probabilities as described in the Equation 3 using a training dataset. To be more specific, the Algorithm 1 pseudo code shows the calculation of the conditional probability.

**Algorithm 1** Calculating $P(C_k|V)$ - conditional probability of class $C_k$ with $V$

```
Sample structure with class label and attribute values
S ← array of training Samples
V ← array of attribute values
C_k ← class number
P_C_k = 0
function get_P_C_k(ClassNumber, Samples)
    output = 0
    for all sample from Samples do
        if sample.getClass() == ClassNumber then
            output += 1/size[Samples]
        end if
    end for
    return output
end function
function get_P_C_k_v_i(ClassNumber, v, i, Samples)
    output = 0
    for all sample from Samples do
        if sample.getClass() == ClassNumber AND sample.getAttribute(i) == v then
            output += 1/size[Samples]
        end if
    end for
    return output
end function
P(C_k|V = 0)
prod = 1
i=0
for all v fromV do
    prod *= get_P_C_k_v_i(ClassNumber,v,i,Samples) / get_P_C_k(ClassNumber,Samples)
    i += 1
end for
P(C_k|V) = prod * get_P_C_k(ClassNumber, Samples))
```

So, we can see from the Equation 1 that given *output* is as a probability that a questioned software sample belongs to one or another class. Therefore, the classification decision will be made by finding a maximal value from set of corresponding class likelihoods. Equation 4 provides formula that assigns class label to the *output*:

$$\hat{y} = \underset{k \in \{1,\ldots,K\}}{\operatorname{argmax}} \; P(C_k)P(V|C_k). \tag{4}$$

**Bayesian Networks** is a probabilistic directed acyclic graphical model (sometimes also named as Bayesian Belief Networks), which shows conditional dependencies using directed acyclic graph. Network can be used to detect *"update knowledge of the state of a subset of variables when other variables (the evidence variables) are observed"* [32]. Bayesian Networks are used in many cases of classification and information retrieval (such as semantic search). The method's routine can be described as following. If edge goes from vertex *A* to vertex *B*, then A is a parent of B, and B is an ancestor of A. If from *A* there is oriented path to another vertex $B$ exists then $B$ is ancestor of $A$, and $A$ is a predecessor of $B$. Lets designate set of parent vertexes of vertex $V_i$ as $parents(V_i) = PA_i$. Direct acyclic graph is called Bayesian Network for probability distribution $P(v)$ given for set of random variables $V$ if each vertex of graph has matched with random variable from $V$. And edge of a graph fits next condition: every variable $v_i$ from $V$ must be conditionally independent from all vertexes that are not its ancestors if all its direct parents $PA_i$ are initialized in graph $G$:

$$\forall V_i \in V \Rightarrow P(v_i|pa_i, s) = P(v_i|pa_i) \tag{5}$$

where $v_i$ is a value of $V_i$, $S$ - set of all vertexes that are not ancestors of $V_i$, $s$ - configuration of $S$, $pa_i$ - configuration of $PA_i$. Then full general distribution of the values in vertexes could be written as product of local distributions, similarly to Naive Bayes rules:

$$\mathrm{P}(V_1, \ldots, V_n) = \prod_{i=1}^{n} \mathrm{P}(V_i \mid \mathrm{parents}(V_i)) \tag{6}$$

Bayesian Belief Networks can be used for classification [32], thus can be applied for malware detection and classification as well [58]. To make Bayesian Network capable of classification it should contain classes as parent nodes which don't have parents themselves. Figure 2 shows an example of such Bayesian network.

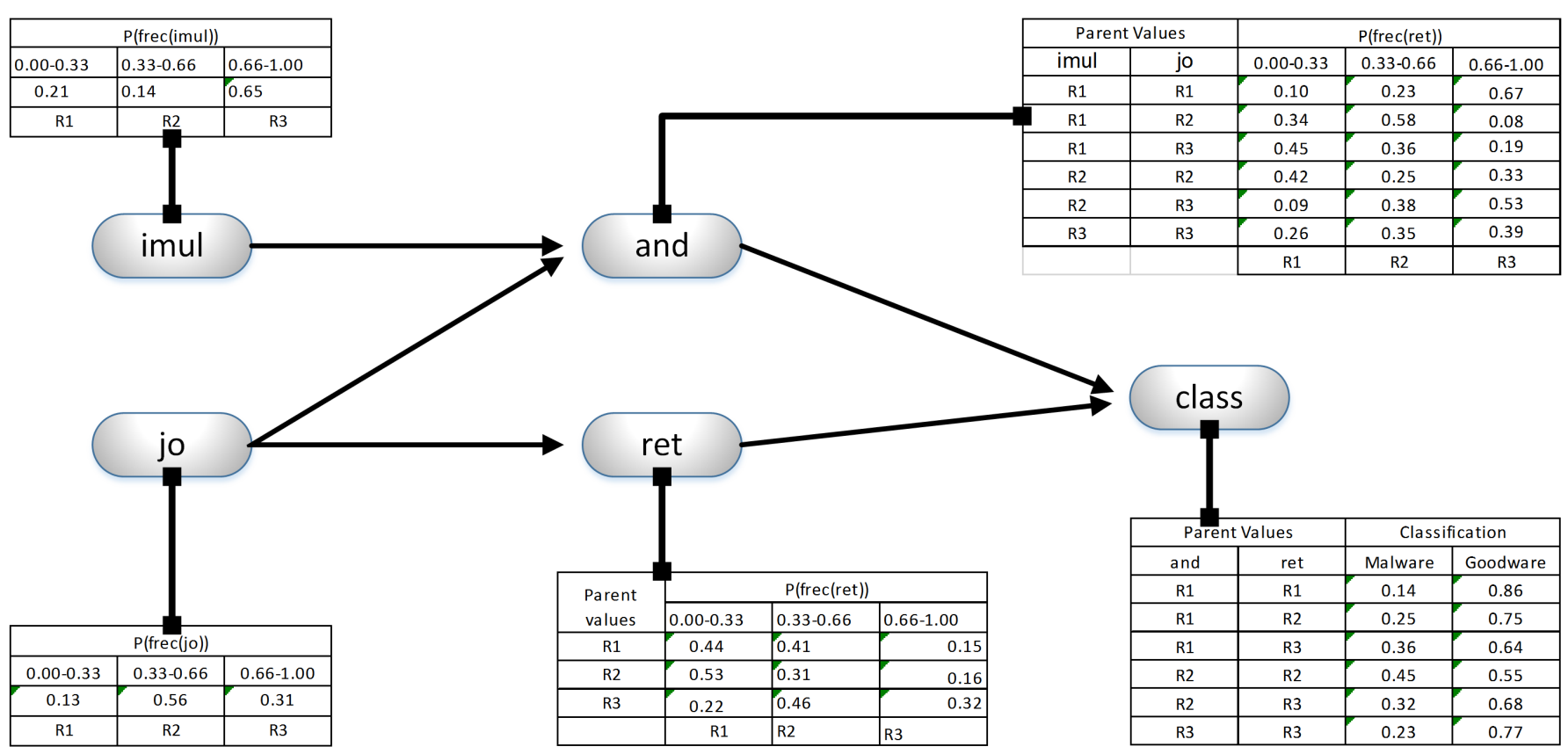

Fig. 2: Bayesian network suitable for malware classification [58]

### 2.2.2 Rule based

Rule based algorithms are used for generating crisp or inexact rules in different Machine Learning approaches [32]. The main advantage of having logic rules involved in malware classification is that logical rules that operate with statements like *equal*, *grater then*, *less or equal to* can be executed on the hardware level which significantly increases the speed of decision making.

**C4.5** is specially proposed by Quinlan [50] to construct decision trees. These trees can be used for classification and especially for malware detection [75]. The process of trees training includes processing of previously classified dataset and on each step looks for an attribute that divides set of samples into subsets with the best information gain. C4.5 has several benefits in compare with other decision tree building algorithms:

- Works not only with discrete but with continuous attributes as well. For continuous attributes it creates threshold tp compares values against [49].
- Take into account missing attributes values.
- Works with attributes with different costs.
- Perform automate tree pruning by going backward through the tree and removing useless branches with leaf nodes.

Algorithm 2 shows a simplified version of decision tree building algorithm.

**Algorithm 2** Decision tree making algorithm

1: $S = s_1, s_2, ...$ labelled training dataset of classified data
2: $x_{1i}, x_{2i}, ..., x_{pi}$ - p-dimensional vector of attributes of each sample $s_i$ form $S$
3: Check for base cases
4: **for all** attributes $x$ **do**
5: Find the normalized gain ratio from splitting set of sample on $x$
6: **end for**
7: Let $x_{best}$ be the attribute with the highest normalized information gain.
8: Create decision *node* that splits on $x_{best}$
9: Repeat on the subsets created by splitting with $x_{best}$. Newly gained nodes add as *children* of current *node*.

**Neuro-Fuzzy** is a hybrid models that ensembles neural networks and fuzzy logic to create human-like linguistic rules using the power of neural networks. Neural network also known as artificial neural network is a network of simple elements which are based on the model of perceptron [65]. Perceptron implements previously chosen activation functions which take input signals and their weights and produces an output, usually in the range of $[0,1]$ or $[-1,1]$. The network can be trained to perform classification of complex and high-dimensional data. Neural Networks are widely used for classification and pattern recognition tasks, thus for malware analysis [72]. The problem is that solutions gained by Neural Networks are usually impossible to interpret because of complexity of internal structure and increased weights on the edges. This stimulates usage of Fuzzy Logic techniques, where generated rules are made in human-like easy-interpretable format: $IF\ \ X > 3\ \ AND\ \ X < 5\ \ THEN\ \ Y = 7$.

Basic idea of Neuro-Fuzzy (NF) model is a fuzzy system that is trained with a learning algorithm similar to one from neural networks theory. NF system can be represented as a neural network which takes input variables and produces output variables while connection weights are represented as encoded fuzzy sets. Thus at any stage (like prior to, in process of and after training) NF can be represented as a set of fuzzy rules. Self-Organising (Kohonen) maps [30] is the most common techniques of combining Neuro and Fuzzy approaches. Shalaginov et al. [64] showed the possibility of malware detection using specially-tuned Neuro-Fuzzy technique on a small dataset. Further, NF showed good performance on large-scale binary problem of network traffic analysis [63]. This method has also proven its efficiency on a set of multinomial classification problems. In particular, it is useful when we are talking about distinguishing not only "malware" or "goodware" but also detecting specific type of "malware" [68]. Therefore, it has been improved for the multinational classification of malware types and families by Shalaginov et al. [70].

### 2.2.3 Distance based

This set of methods is used for classification based on predefined distance measure. Data for distance-based methods should be carefully prepared, because computational complexity grows significantly with space dimensionality (number of features) and number of training samples. Thus there is a need for proper feature selection as well as sometimes for data normalization.

**k-Nearest Neighbours** or k-NN is classification and regression method. k-NN does not need special preparation of the dataset or actual "training" as the algorithm is ready for classification right after labelling the dataset. The algorithm takes a sample that is need to be classified and calculates distances to samples from training dataset, then it selects $k$ nearest neighbours (with shortest distances) and makes decision based on class of this $k$ nearest neighbours. Sometimes it makes decision just on the majority of classes in this $k$ neighbours selection, while in other cases there is weights involved in process of making decision. When k-NN is used for malware classification and detection there is a need for careful feature selection as well as a methodology for dealing with outliers and highly mixed data, when training samples cannot create distinguishable clusters [58].

**Support Vector Machine** or SVM is a supervised learning method. It constructs one or several hyperplanes to divide dataset for classification. Hyperplane is constructed to maximize distance from it to the nearest data points. Sometimes kernel transformation is used to simplify hyperplanes. Building a hyperplane is usually turned into two-class problem (one vs one, one vs many) and involves quadratic programming. Lets have linearly separable data (as shown in Figure 3) which can be represented as $\mathscr{D} = \{(\mathbf{x}_i, y_i) \mid \mathbf{x}_i \in \mathbb{R}^p, y_i \in \{-1, 1\}\}_{i=1}^{n}$. Where $y_i$ is 1 or -1 depending on class of point $x_i$. Each $x_i$ is p-dimensional vector (not always normalised). The task is to find hyperplane with maximum margin that divides dataset on points with $y_i = 1$ and $y_i = -1$: $w \cdot x - b = 0$. Where $w$ is a normal vector to a hyperplane [21]. If dataset is linearly separable we can build two hyperplanes $w \cdot x - b = 1$ and $w \cdot x - b = -1$ between which there will be no (or in case of soft margin maximal allowed number) points. Distance between them (margin) is $\frac{2}{||w||}$, so to maximize margin we need to minimize $||w||$ and to find parameters of hyperplane we need to introduce Lagrangian multipliers $\alpha$ and solve Equation 7 with quadratic programming techniques.

$$\arg\min_{\mathbf{w},b}\max_{\alpha\geq 0}\left\{\frac{1}{2}\|\mathbf{w}\|^2-\sum_{i=1}^{n}\alpha_i[y_i(\mathbf{w}\cdot\mathbf{x_i}-b)-1]\right\} \tag{7}$$

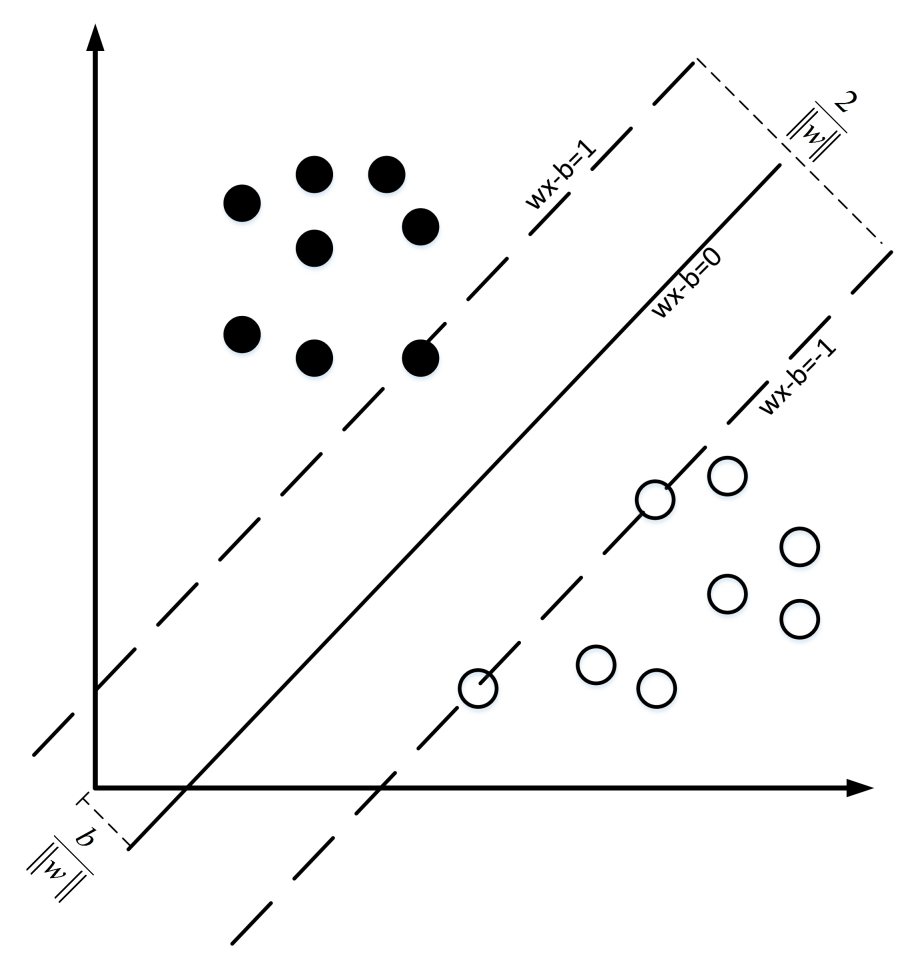

Fig. 3: Maximum margin hyperplane for two class problem [32]

Sometimes there is a need to allow an algorithm to work with misclassified data hence leaving some points inside the margin based on the degree of misclassification $\xi$. So the Equation 3 turns into Equation 8.

$$\arg\min_{\mathbf{w},\xi,b}\max_{\alpha,\beta}\left\{\frac{1}{2}\|\mathbf{w}\|^2+C\sum_{i=1}^{n}\xi_i-\sum_{i=1}^{n}\alpha_i[y_i(\mathbf{w}\cdot\mathbf{x_i}-b)-1+\xi_i]-\sum_{i=1}^{n}\beta_i\xi_i\right\} \tag{8}$$
$$with\ \ \alpha_i,\beta_i\geq 0$$

Also the data might be linearly separated, so there is a need for kernel trick. The basic idea is to substitute every *dot product* with non-linear kernel function. Kernel function can be chosen depending on situation and can be polynomial, Gaussian, hyperbolic etc. SVM is a very powerful technique which can give good accuracy if properly used, so it often used in malware detection studies as shown by Ye et al. [82].

### 2.2.4 Neural networks

Neural Network is based on the model of perceptron which has predefined activation function. In the process of training weights of the links between neurons are trained to fit train data set with minimum error with use of back propagation. Artificial Neural network (ANN) consists of input layer, hidden layer (layers) and output layer as it is shown on Figure 4.

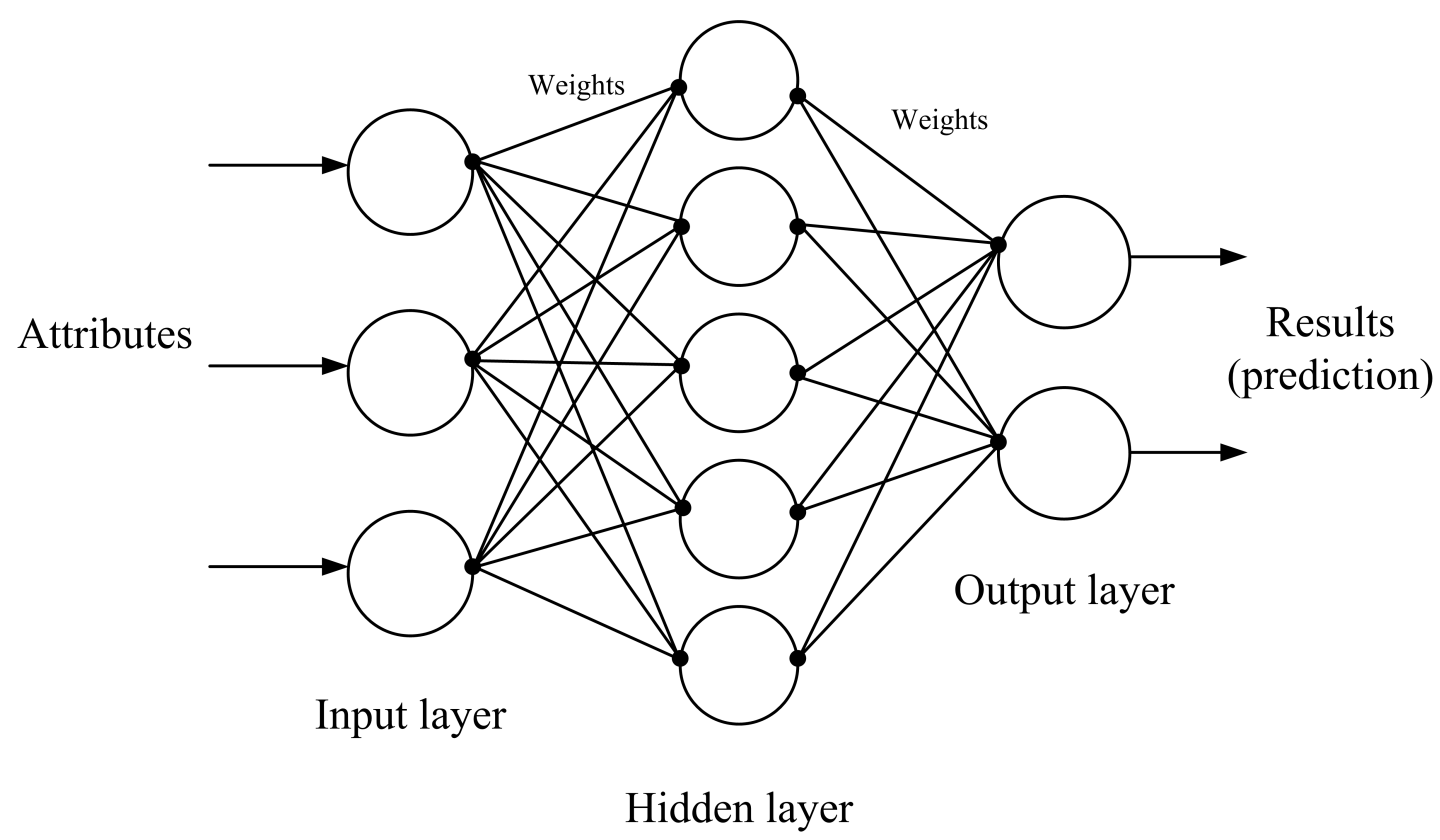

Fig. 4: Artificial neural network [32]

The input layer takes normalized data, while hidden output layer produces activation output using neuron's weighted input and activation function. Activation function is a basic property of neuron that takes input values given on the input edges, multiply them by weights of these edges and produces output usually in a range of [0,1] or [-1,1]. Output layer is needed to present results and then interpret them. Training of ANN starts with random initialization of weights for all edges. Then feature vector of each sample is used as an input. Afterwards, result gained on the output layer is compared to the real answers. Any errors are calculated and using back-propagation all weights are tuned. Training can continue until reaching desired number of training cycles or accuracy. Learning process of ANN can be presented as shown in the Algorithm 3. Artificial Neural networks can be applied for complex models in high-dimensional spaces. This is why it often used for malware research [72].

**Algorithm 3** ANN training

```
S = s_1, s_2, ... labeled training dataset of classified data
x_1i, x_2i, ..., x_pi - p-dimensional vector of attributes of each sample s_i form S
N number of training cycles
L_rate learning rate
Random weight initialization
for all training cycles N do
    for all samples S do
        give features x_i as input to the ANN
        compare class of s_i with gained output of ANN
        calculate error
        using back-propagation tune weights inside the ANN with L_rate
    end for
    reduce L_rate
end for
```

#### 2.2.5 Open Source and Freely available ML Tools

Today machine learning is widely used in many areas of research with many publicly available tools (Software products, libraries etc.).

**Weka** or Waikato Environment for Knowledge Analysis is a popular, free, cross platform and open source tool for machine learning. It supports many of popular ML methods with possibility of fine tuning of the parameters and final results analysis. It provides many features such as splitting dataset and graphical representation of the results. Weka results are saved in .arff file which is specially prepared CSV file with header. It suffers from couple of issues including no support for multi-thread computations and poor memory utilization especially with big datasets.

**Python weka wrapper** is the package which allows using power of Weka through Python programs. It uses javabridge to link Java-based Weka libraries to python. It provides the same functionality as Weka, but provides more automation capacities.

**LIBSVM** and LIBLINEAR are open source ML libraries written in C++ supporting kernelized SVMs for linear, classification and regression analysis. Bindings for Java, Mathlab and R are also present. It uses space-separated files as input, where zero values need not to be mentioned.

**RapidMiner** is machine learning and data mining tool with a user friendly GUI and support for a lot of ML and data mining algorithms.

**Dlib** is a free and cross-platform C++ toolkit which supports different -machine learning algorithms and allows multi-threading and utilization of Python APIs.

#### 2.2.6 Feature Selection & Construction process

Next important step after the characteristics extraction is so-called Feature Selection process [32]. Feature Selection is a set of methods that focus on elimination of irrelevant or redundant features that are not influential for malware classification.

This is important since the number of characteristics can be extremely large, while only a few can actually be used to differentiate malware and benign applications with a high degree of confidence. The most common feature selection methods are *Information Gain*, and *Correlation-based Feature Subset Selection (CFS)* [24]. The final goal of Feature Selection is to simplify the process of knowledge transfer from data to a reusable classification model.

## *2.3 Taxonomy of malware static analysis using Machine Learning*

Our extensive literature study as reflected in Table 1 resulted to proposing a taxonomy for malware static analysis using machine learning as shown in Figure 5. Our taxonomy depicts the most common methods for analysis of static characteristics, extracting and selecting features and utilizing machine learning classification techniques. Statistical Pattern Recognition process [26] was used as the basis for our taxonomy modelling.

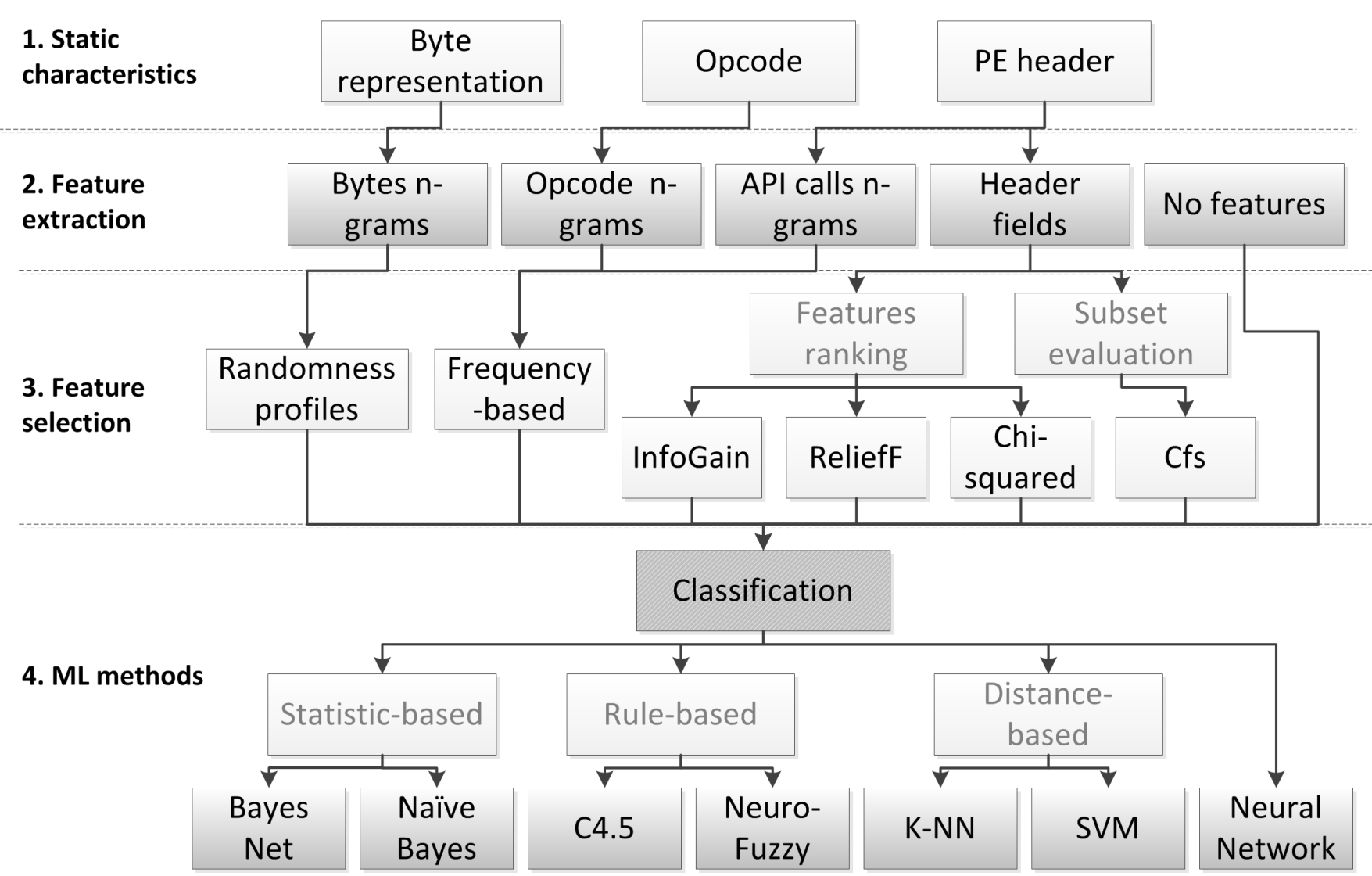

Fig. 5: Taxonomy of common malware detection process based on static characteristics and Machine Learning

| Year | Authors | Dataset | Features | FS | ML |
|---|---|---|---|---|---|
| | | | *PE32 header* | | |
| 2016 | Cepeda et al. [7] | 7,630 malware and 1.818 goodware | 57 features from VirisTotal | ChiSqSelector with 9 features finally | SVM, RF, NN |
| 2016 | Le-Khac et al. [35] | Malicious: 94 ;Benign: 620 | Control Flow Change and 2-6 n-grams | - | Naive Bayes |
| 2014 | Markel et al. [38] | Malicious: 122,799, Benign: 42,003 | 46 features use python 'pefile' | - | Naive Bayes, Logistic Regression, Classification and Regression Tree (CART) |
| 2013 | Khorsand et al. [29] | Benign: 850 EXE and 750 DLL; Malware: 1600 from VX heavens | eliminated | - | Prediction by partial matching |
| 2012 | Devi et al. [15] | 4,075 PE files: 2954 malicious and 1121 Windows XP SP2 benign | 2 + 5 features | - | BayesNet, k-NN, SVM, AdaBoostM1, Decision table, C4.5, Random Forest, Random Tree |
| 2011 | Zhao [84] | 3109 PE: 1037 viruses from Vx Heavens and 2072 benign executable on Win XP Sp3 | 24 features from PE files using Control Flow Graph-based on nodes | | Random Forest, Decision Tree, Bagging, C4.5 |
| 2011 | Ugarte-Pedrero et al. [77] | 500 benign from WinXP and 500 non-packed from Vxheaven; 500 packed + 500 Zeus | 166 structure features of PE file | InfoGain | Learning with Local and Global Consistency, Random Forest |
| 2011 | Santos et al. [59] | 500 benign and 500 malicious from VxHeavens, also packed and not packed | 209 structural features | InfoGain | Collective Forest |
| 2009 | Tang [76] | 361 executables and 449 normal trojan files | PE header structural features | - | Decision Tree |
| 2009 | Wang et al. [80] | Benign: 1,908, Malicious: 7,863 | 143 PE header entries | InfoGain, Gain raio | SVM |
| | | | *bytes n-gram sequences* | | |
| 2011 | Jain et al. [27] | 1,018 malware and 1,120 benign samples | 1-8 byte, n-gram, best n-gram by documentwise frequency | - | NB, iBK, J48, AdaBoost1, RandomForest |
| 2007 | Masud et al. [39] | 1st set - 1,435 executables: 597 of which are benign and 838 are malicious. 2nd set - 2,452 executables: 1,370 benign and 1,082 malicious | 500 best n-grams | InfoGain | SVM |
| 2006 | Reddy et al. [51] | 250 malware vs 250 benign | 100-500 best n-gram | Document Frequency, InfoGain | NB, iBK, Decision Tree |
| 2004 | Kolter et al. [31] | 1971 benign, 1651 malicious from Vx Heaven | 500 best n-grams | InfoGain | Naive Bayes, SVM, C4.5 |
| | | | *opcode n-gram sequences* | | |
| 2016 | Wang et al. [79] | 11,665 malware and 1,000 benign samples | 2-tuple opcode sequences | information entropy | density clustering |
| 2015 | Bragen [6] | 992 malwares, 771 benign from Windows Vista | 1-4 n-gram opcode with vocabulary 530-714,390 | Cfs, Chi-sqaured, InfoGain, ReliefF, SymUncert. | Random Forest, C4.5, Naive Bayes, bayes Net, Baggin, ANN, SOM, k-nn |
| 2013 | Santos et al. [58] | 13,189 malware vs 13,000 benign | top 1,000 features | InfoGain | Random Forest, J48, k-Nearest Neighbours, Bayesian networks, SVM |
| 2011 | Shahzad et al. [62] | Benign: 300, Malicious: 300 on Windows XP | coabulary of 1,413 with n-gram=4 | tf-idf | ZeroR, Ripper, C4.5, SVM, Naive Bayes, k-nn |
| | | | *API calls* | | |
| 2012 | Zabidi et al. [83] | 23 malware and 1 benign | API calls, debugger features, VM features | - | - |
| 2012 | Faruki et al. [19] | 3234 benign, 3256 malware | 1-4 API call-gram | - | Random Forest, SVM, ANN, C4.5, Naive bayes |
| 2010 | Shankarapani et al. [71] | 1593 PE files:875 benign and 715 malicious | API calls sequence | - | SVM |
| 2010 | Sami et al. [55] | 34,820 PE: 31,869 malicious and 2951 benign from Windows | API calls | Fisher Score | Random Forest, C4.5, Naive Bayes |
| | | | *no features / not described* | | |
| 2012 | Baig et al. [5] | 200 packed PE and 200 unpacked from Windows 7, Windows 2003 Server | file entropy | - | - |
| 2010 | Dube et al. [17] | 40,498 samples: 25,974 malware, 14,524 benign | from 32 bit files | - | Decision Tree |

Table 1: Analysis of ML methods applicability for different types of static characteristics

To get a clear picture on application domain of each machine learning and feature selection method we analysed reported performance as shown in Figure 6.Majority of researchers were using byte n-gram, opcode n-gram and PE32 header fields for static analysis while C4.5, SVM or k-NN methods were mainly used for malware detection. Information Gain is the prevalent method to define malware attributes. Also we can see that n-gram-based method tend to use corresponding set of feature selection like tf-idf and Symmetric Uncertainty that are more relevant for large number of similar sequences. On the other hand, PE32 header-based features tend to provide higher entropies for classification and therefore Control-Flow graph-based and Gain Ratio are more suitable for this task.

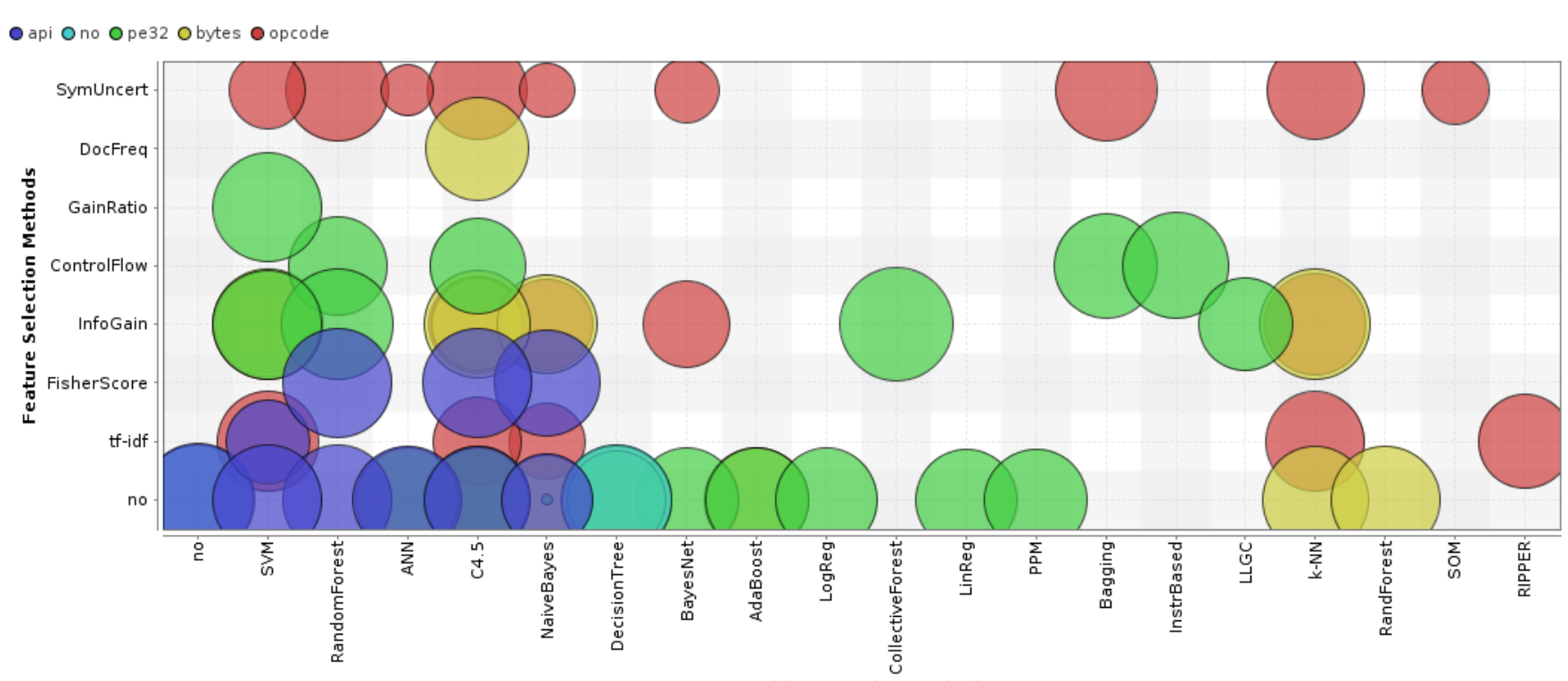

Fig. 6: Comparison of accuracy of various static characteristics with respect to feature selection and machine learning methods. Colour of the bubbles shows used characteristics for detection, while size of the bubble denotes achieved accuracy

To conclude, one can say that majority of authors either extract features that offers good classification accuracy, or use conventional methods like Information Gain. However, n-gram based characteristics need other FS approaches to eliminate irrelevant features. Rule-based ML is the most commonly used classification method along with SVM. Forest-based method tends to be more applicable for PE32 header-based features. Also ANN is not commonly-used technique. While most of the works achieved accuracies of 80-100%, some Bayes-based methods offered much lower accuracy even down to 50% only.

## 3 Approaches for Malware Feature Construction

Similar the previous works, following four sets of static properties are suggested for feature classification in this paper:

**PE32 header** features characterize the PE32 header information using the *PEframe* tools [77]. Following numerical features will be used in our experiments:

- *ShortInfo_Directories* describes 16 possible data directories available in PE file. The most commonly used are "Import", "Export", "Resource", "Debug", "Relocation".
- *ShortInfo_Xor* indicates detected XOR obfuscation.
- *ShortInfo_DLL* is a binary flag of whether a file is executable or dynamically-linked library.
- *ShortInfo_FileSize* measures size of a binary file in bytes.
- *ShortInfo_Detected* shows present techniques used to evade the detection by anti-viruses like hooks to disable execution in virtualized environment or suspicious API calls.
- *ShortInfo_Sections* is a number of subsections available in the header.
- *DigitalSignature* contains information about the digital signatture that can be present in a file
- *Packer* describes used packer detected by *PEframe*
- *AntiDebug* gives insight into the techniques used to prevent debugging process.
- *AntiVM* is included to prevent the execution in a virtualize environment.
- *SuspiciousAPI* indicates functions calls that are labelled by *PEframe* as suspicious.
- *SuspiciousSections* contains information about suspicious sections like ".rsrc \u0000 \u0000 \u0000"
- *Url* is a number of different url addresses found in the binary file.

**Byte n-gram**. N-gram is a sequence of some items (with minimum length of 1) that are predefined as minimal parts of the object expressed in Bytes. By having the file represented as a sequence of bytes we can construct 1-gram, 2-gram, 3-gram etc. N-grams of bytes, or byte n-grams are widely used as features for machine learning and static malware analysis [27, 51].

Reddy et al. [51] used n-grams of size 2,3 and 4 with combination of SVM, Instance-based learner and Decision Tree algorithms to distinguish between malicious and benign executables. After extracting n-grams they used class-wise document frequency as a feature selection measure and showed that class-wise document frequency is performing better than Information Gain as a feature selection measure. Jain et al. [27] used n-grams in range of 1 to 8 as features and Naive Bayes, Instance Based Learner and AdaBoost1 [3] as machine learning algorithms for malware classification and reported byte 3-grams as the best technique.

**Opcode n-gram** represent a set of instructions that will be performed on the CPU when binary is executed. These instructions are called operational codes or **opcodes**. To extract opcodes from executable we need to perform disassembly procedure. After this opcodes will be represented as short instructions names such as *POP, PUSH, MOV, ADD, SUB* etc. Santos et al. [58] described a method to distinguish between malicious and benign executables or detecting different malware families using opcode sequences of length 1 to 4 using Random Forest, J48, k-Nearest Neighbours, Bayesian Networks and Support Vector Machine algorithms [3].

**API calls** is a set of tools and routines that help to develop a program using existent functionality of an operating system. Since most of the malware samples are platform dependent it is very much likely that their developers have use APIs as well. Therefore, analysing API calls usage among benign and malicious software can help to find malware-specific API calls and therefore are suitable to be used as a feature for machine learning algorithms. For example, [71] successfully used Support Vector Machines with frequency of API calls for malware classification. [78] provided a methodology for classification of malicious and benign executables using API calls and n-grams with n from 1 to 4 and achieved accuracy of 97.23% for 1-gram features. [19] used so-called API call-gram model with sequence length ranging from 1-4 and reached accuracy of 97.7% was achieved by training with 3-grams. In our experiments we are going to use 1 and 2 n-grams as features generated from API calls.

## 4 Experimental Design

All experiments were conducted on a dedicated Virtual machine (VM) on Ubuntu 14.04 server running on Xen 4.4. The server had an Intel(R) Core(TM) i7-3820 CPU @ 3.60GHz with 4 cores (8 threads), out of which 2 cores (4 threads) were provided to the VM. Disk space is allocated on the SSD RAID storage based on Samsung 845DC. Installed server memory was Kingston PC-1600 RAM, out of which 8GB was available for the VM. Operating system was an Ubuntu 14.04 64 bit running on ae dedicated VM together with all default tools and utilities available in the OS's repository. Files pre-processing were performed using *bash* scripts due to native support in Linux OS. To store extracted features we utilised MySQL 5.5 database engine together with Python v 2.7.6 and PHP v 5.5.9 connectors.

For the experiments we used a set of benign and malicious samples. To authors knowledge there have not been published any large BENIGN SOFTWARE REFERENCE DATASETS, so we have to create our own set of benign files. Since the focus of the paper is mainly on PE32 Windows executables, we decided to extract corresponding known-to-be-good files from different versions of MS Windows, including different software and multimedia programs installations that are available. The OSes that we processed were 32 bit versions of Windows XP, Windows 7, Windows 8.1 and Windows 10. Following two Windows malware datasets were used in our research:

1. VX HEAVEN [2] dedicated to distribute information about the computer viruses and contains 271,092 sorted samples dating back from 1999.
2. VIRUS SHARE [1] represent sharing resource that offers 29,119,178 malware samples and accessible through VirusShare tracker as of 12th of July, 2017. We utilized following two archives: *VirusShare_00000.zip* created on 2012-06-15 00:39:38 with a size of 13.56 GB and *VirusShare_00207.zip* created on 2015-12-16 22:56:17 with a size of 13.91 GB, all together contained 131,072 unique, uncategorised and unsorted malware samples. They will be referred further as *malware_000* and *malware_207*.

To be able to perform experiments on the dataset, we have to filter out irrelevant samples (not specific PE32 and not executables), which are out of scope in this paper. However, processing of more than 100k samples put limitations and require non-trivial approaches to handle such amount of files. We discovered that common ways of working with files in directory such that simple *ls* and *mv* in *bash* take unreasonable amount of time to execute. Also there is no way to distinguish files by extension like **.dll* or **.exe* since the names are just *md5* sums. So, following filtering steps were performed:

1. Heap of unfiltered malware and benign files were placed into two directories ”malware/” and ”benign/”.
2. To eliminate duplicates, we renamed all the files to their MD5 sums.
3. PE32 files were detected in each folder using *file* Linux command:

```
\$ file 000000b4dccfbaa5bd981af2c1bbf59a
000000b4dccfbaa5bd981af2c1bbf59a: PE32 executable (
    DLL) (GUI) Intel 80386, for MS Windows
```

4. All PE32 files from current directory that meet our requirement were scrapped and move to a dedicated one:

```
#!/bin/sh
cd ../windows1;
counter=0;
for i in *; do
counter=$((counter+1));
echo "$counter";
VAR="file $i | grep PE32 " ;
```

```
 VAR1=$(eval "$VAR") ;
 len1=${#VAR1};
if [ -n "$VAR1" ] && [ "$len1" -gt "1" ] ;
then
 echo "$VAR1" | awk '{print $1}' | awk '{gsub(/:$/,""
    ); print $1 " ../windows/PE/" $1}'| xargs mv -f ;
else
 echo "other";
 file $i | awk '{print $1}' | awk '{gsub(/:$/,"");
    print $1 " ../windows/other/" $1}' | xargs mv -f
    ;
 fi
done
```

5. We further can see a variety of PE32 modifications for 32bit architecture:

   ```
   PE32 executable (GUI) Intel 80386, for MS Windows
   PE32 executable (DLL) (GUI) Intel 80386, for MS
       Windows
   PE32 executable (GUI) Intel 80386, for MS Windows,
       UPX compressed
   ```

   Following our purpose to concentrate on 32bit architecture, only PE32 are filtered out from all possible variants of PE32 files shown about.
6. After extracting a target group of benign and malicious PE32 files, multiple rounds of feature extraction are performed according to methods used in the literature.
7. Finally, we insert extracted features into the corresponding MySQL database to ease the handling, feature selection and machine learning processes respectively.

After collecting all possible files and performing the pre-processing phase, we ended up with the sets represented in the Table 2.

Table 2: Characteristics of the dataset collected and used for our experiments after filtering PE files

| **Dataset** | **Number of files** | **Size** |
|---|---|---|
| Benign | 16,632 | 7.4GB |
| Malware_000 | 58,023 | 14.0GB |
| Malware_207 | 41,899 | 16.0GB |

Further, feature construction and extraction routine from PE files was performed using several tools as follows:

1. PEFRAME [4] is an open source tool specifically designed for static analysis of PE malware. It extracts various information from PE header ranging from packers to anti debug and anti vm tricks.

2. HEXDUMP is a standard Linux command line tool which is used to display a file in specific format like ASCII or one-byte octal.
3. OBJDUMP is a standard Linux command line too to detect applications instructions, consumed memory addresses, etc.

## 5 Results & Discussions

Before testing different ML techniques for malware detection it is important to show that our datasets actually represent the real-world distribution of the malware and goodware. Comparison of "Compile Time" field of PE32 header can be utilized for this purpose. Figure 7 represents log-scale histogram of the compilation time for our benign dataset. Taking into consideration the Windows OS timeline we found a harmony between our benign dataset applications compile time and development of Microsoft Windows operating systems. To start with, **Windows 3.1** was originally released on April 6, 1992 and our plot of benign applications indicates the biggest spike in early 1992. Later on in 1990th, **Windows 95** was due on 24 August 1995, while next **Windows 98** was announced on 25 June 1998. Further, 2000th marked release of **Windows XP** on October 25, 2001. Next phases on the plot correspond to the release of **Windows Vista** on 30th January 2007 and **Windows 7** on 22nd October 2009. Next popular version (Windows 8) appeared on 26th October 2012 and the latest major spike in the end of 2014 corresponds to the release of **Windows 10** on 29th July 2015.

Further, compilation time distribution for the first malware dataset *malware_000* is given in the Figure 8. We can clearly see that release of newer Windows version is always followed by an increase of cumulative distribution of malware samples in following 6 to 12 month. It can be seen that the release of 32bit Windows 3.1 cause a spike in a number of malware. After this the number of malware compiled each year is constantly growing. Then, another increase can be observed in second half of the 2001 which corresponds to the release of the Windows XP and so on.

Considering the fact that MS DOS was released in 1981 it makes compilation times before this day look like fake or just obfuscated intentionally. On the other hand the dataset *malware_000* cannot have dates later than June 2012. Therefore, malware with compilation time prior to 1981 or later than Jan 2012 are tampered

### *5.1 Accuracy of ML-aided Malware Detection using Static Characteristics*

This part presents results of apply Naive Bayes, BayesNet, C4.5, k-NN, SVM, ANN and NF machine learning algorithms against static features of our dataset namely PE32 header, Bytes n-gram, Opcode n-gram, and API calls n-gram.

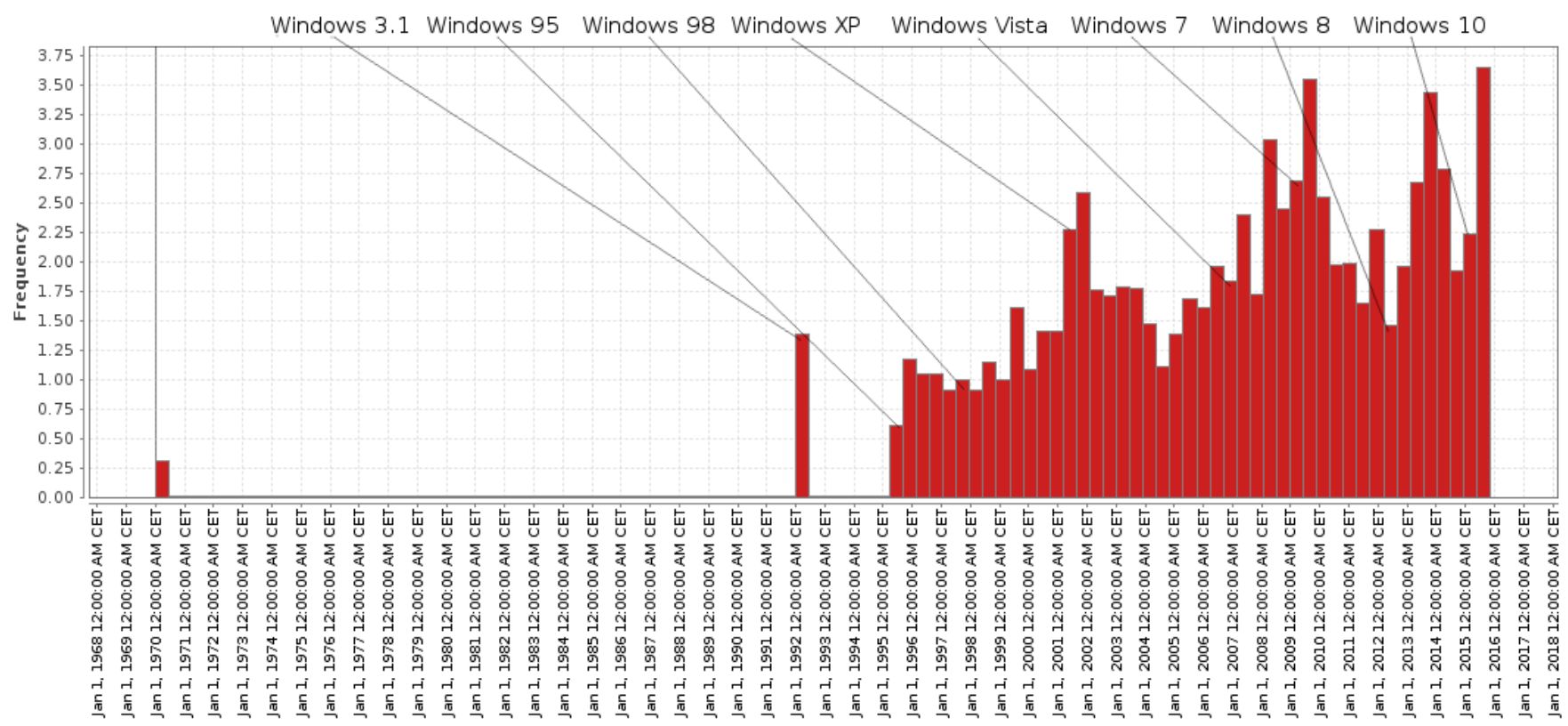

Fig. 7: Log-scale histogram of compilation times for *benign* dataset

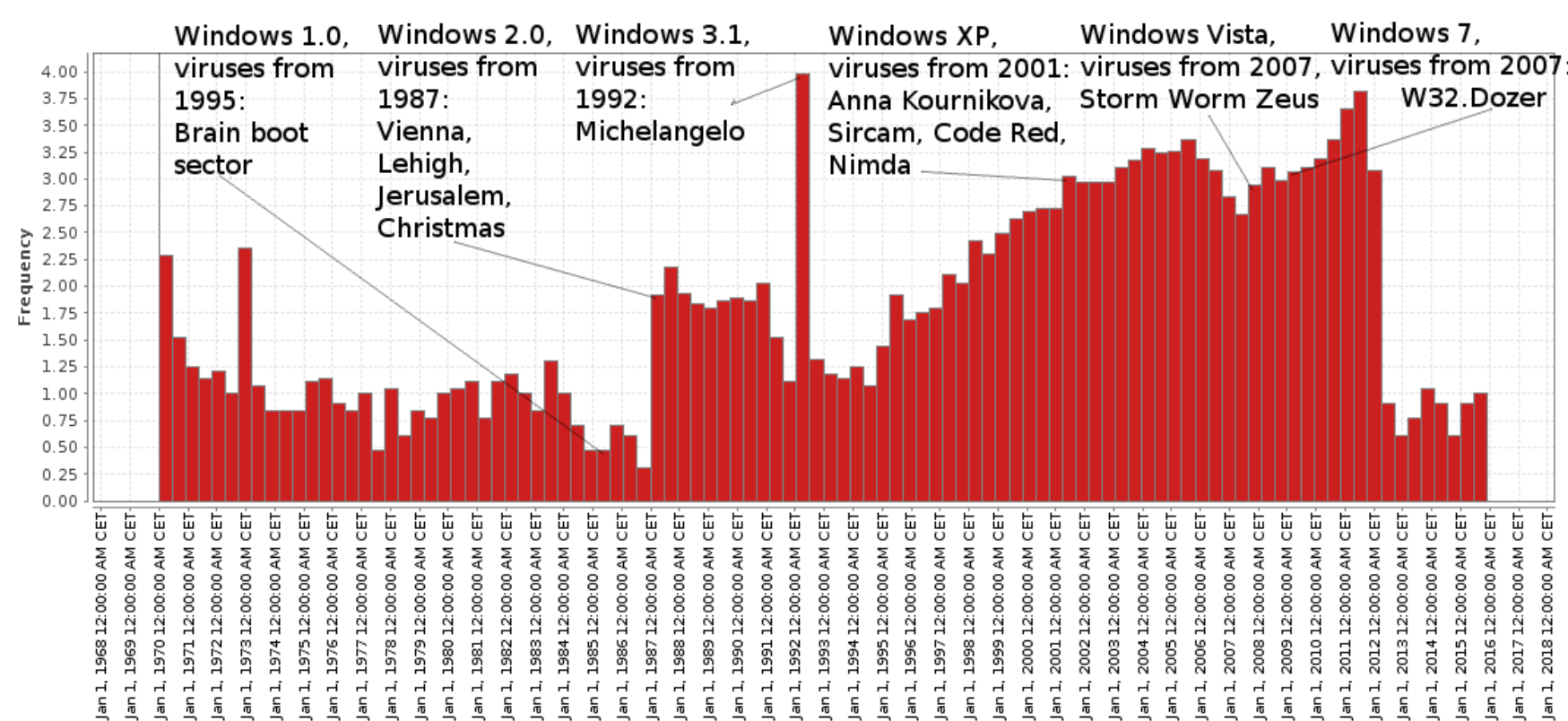

Fig. 8: Log-scale histogram of compilation times for *malware_000* dataset

### 5.1.1 PE32 header

PE32 header is one of the most important features relevant to threat intelligence of PE32 applications. We performed feature selection using *Cfs* and *InfoGain* methods with 5-fold cross-validation as presented in the Table 3.

We can clearly see that the features from the *Short Info* section in PE32 headers can be used as a stand-alone malware indicators, including different epochs. Number of directories in this section as well as file size and flag of EXE or DLL have bigger merits in comparison to other features. To contrary, *Anti Debug* and *Suspicious API* sections from *PEframe* cannot classify a binary file. Finally, we can say that digital signature and *Anti VM* files in PE32 headers are almost irrelevant in malware detection. Further, we performed exploration of selected ML methods that can be used with selected features. By extracting corresponding numerical features mentioned earlier, we were able to achieve classification accuracy levels presented in Table 4. Table 3 presents also accuracy of ML method after performing feature selection. Here we used whole sub-sets defined by *Cfs* method and features with merit of $\geq 0.1$ detected by *InfoGain*.

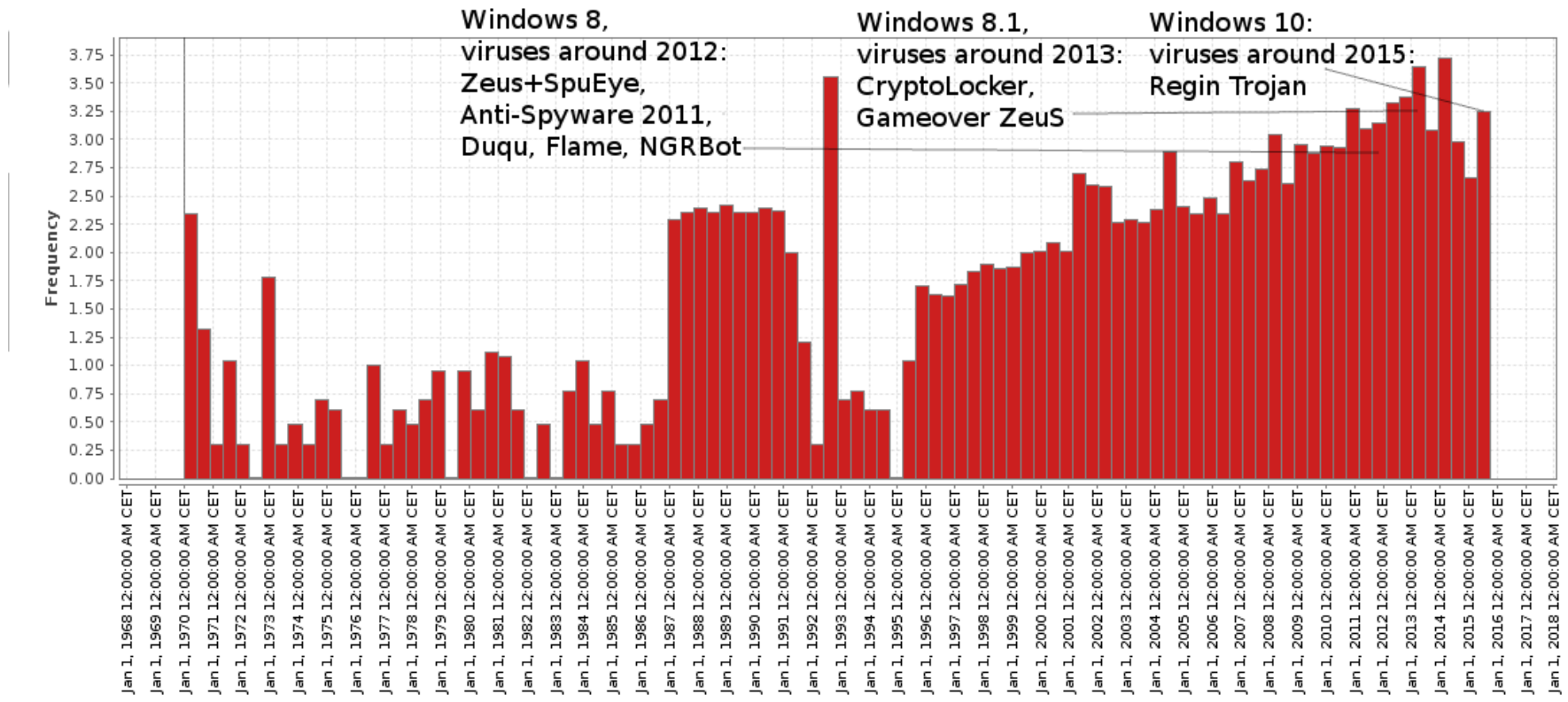

Fig. 9: Log-scale histogram of compilation times for *malware_207* dataset

Table 3: Feature selection on PE32 features. Bold font denotes selected features according to *InfoGain* method

| Benign vs Malware_000 | | Benign vs Malware_207 | | Malware_000 vs Malware_207 | |
|---|---|---|---|---|---|
| **Information Gain** | | | | | |
| merit | attribute | merit | attribute | merit | attribute |
| 0.377 | **ShortInfo_Directories** | 0.369 | **ShortInfo_DLL** | 0.131 | **ShortInfo_FileSize** |
| 0.278 | **ShortInfo_DLL** | 0.252 | **ShortInfo_Directories** | 0.094 | ShortInfo_Detected |
| 0.118 | **AntiDebug** | 0.142 | **ShortInfo_FileSize** | 0.064 | SuspiciousAPI |
| 0.099 | Packer | 0.105 | **SuspiciousSections** | 0.044 | ShortInfo_Directories |
| 0.088 | SuspiciousSections | 0.101 | **SuspiciousAPI** | 0.036 | Packer |
| 0.082 | ShortInfo_Xor | 0.089 | AntiDebug | 0.028 | AntiDebug |
| 0.076 | SuspiciousAPI | 0.084 | ShortInfo_Detected | 0.017 | SuspiciousSections |
| 0.045 | ShortInfo_FileSize | 0.054 | ShortInfo_Xor | 0.016 | Url |
| 0.034 | ShortInfo_Detected | 0.050 | Packer | 0.015 | AntiVM |
| 0.022 | Url | 0.036 | Url | 0.012 | ShortInfo_Xor |
| 0.004 | AntiVM | 0.002 | AntiVM | 0.002 | ShortInfo_DLL |
| 0 | ShortInfo_Sections | 0 | ShortInfo_Sections | 0 | ShortInfo_Sections |
| 0 | DigitalSignature | 0 | DigitalSignature | 0 | DigitalSignature |
| **Cfs** | | | | | |
| attribute | | attribute | | attribute | |
| ShortInfo_Directories | | ShortInfo_Directories | | ShortInfo_Directories | |
| ShortInfo_Xor | | ShortInfo_Xor | | ShortInfo_FileSize | |
| ShortInfo_DLL | | ShortInfo_DLL | | ShortInfo_Detected | |
| ShortInfo_Detected | | | | Packer | |
| Url | | | | | |

Table 4: Comparative classification accuracy based on features from PE32 header, in %. Bn, Ml_000 and Ml_207 are benign and two malaware datasets respectively

| Dataset | Naive Bayes | BayesNet | C4.5 | k-NN | SVM | ANN | NF |
|---|---|---|---|---|---|---|---|
| **All features** | | | | | | | |
| Bn vs Ml_000 | 90.29 | 91.42 | **97.63** | 97.30 | 87.75 | 95.08 | 92.46 |
| Bn vs Ml_207 | 88.27 | 91.21 | **96.43** | 95.99 | 84.88 | 93.24 | 89.03 |
| Ml_000 vs Ml_207 | 63.41 | 71.59 | **82.45** | 82.11 | 73.77 | 69.99 | 69.01 |
| **Information Gain** | | | | | | | |
| Bn vs Ml_000 | 88.32 | 89.17 | **94.09** | 94.01 | 94.09 | 93.51 | 87.53 |
| Bn vs Ml_207 | 87.25 | 90.39 | **95.06** | 94.58 | 84.55 | 92.37 | 87.88 |
| Ml_000 vs Ml_207 | 58.26 | 67.05 | 67.77 | **70.70** | 69.46 | 63.19 | 51.31 |
| **Cfs** | | | | | | | |
| Bn vs Ml_000 | 89.35 | 90.89 | **95.39** | 95.38 | 95.16 | 93.69 | 85.85 |
| Bn vs Ml_207 | 86.88 | 89.67 | 91.61 | **91.68** | 91.68 | 91.68 | 81.91 |
| Ml_000 vs Ml_207 | 67.45 | 70.95 | 76.98 | **76.92** | 72.15 | 68.18 | 67.06 |

Malware and goodware can be easily classified using full set as well as sub-set of features. One can notice that ANN and C4.5 performed much better than other methods. It can be also seen that the high quality of these features made them very appropriate to differentiate between the *benign* and *malware_000* dataset. Further, we can see that the two datasets *malware_000* and *malware_207* are similar and extracted features do not provide a high classification accuracy. Neural Network was used with 3 hidden layers making it a non-linear model and experiments were performed using 5-fold cross-validation technique.

### 5.1.2 Bytes n-gram

Bytes n-gram is a very popular method for static analysis of binary executables. This method has one significant benefit: in order to perform analysis there is no need of previous knowledge about file type and internal structure since we use its raw (binary) form. For feature construction we used random profiles that were first presented by Ebringer et al. [18] called *fixed sample count* (see Figure 10), which generates fixed number of random profiles regardless of the file size and *sliding window algorithm*. In this method each file is represented in a hexadecimal format and frequencies of each byte are counted to build a Huffman tree for each file. Then using window of fixed size and moving it on fixed skip size the randomness profile of each window is calculated. A Randomness profile is sum of Huffman code length of each byte in a single window. The lower the randomness in a particular window the bigger will be the randomness of that profile.

We chose 32 bytes as the most promising sliding window size [18, 48, 74] and due to big variety of file sizes in our dataset, we chose 30 best features (or *pruning size* in terminology from [18]) which are the areas of biggest randomness (the most unique parts) in their original order. This features was fed into different machine

Generate randomness measurements for a file, output proportional to file length

INPUT: An array of bytes $b_1, \ldots, b_n$, a window size $w$ and a skip size $a$.

OUTPUT: An array of $\left\lceil \frac{n-w}{a} \right\rceil$ samples of the randomness, ranging from 0.0 to1.0.

1. Construct the Huffman tree for the input bytes.
2. Construct an array $e_1, \ldots, e_{256}$ containing the encoding length for each of the input bytes.
3. For $i$ from 1 to $\left\lceil \frac{n-w}{a} \right\rceil$, do the following:
   a) $r_i \leftarrow \sum_{j=a(i-1)+1}^{a(i-1)+w+1} e_j$
4. Rescale $r_i$ between 0.0 and 1.0, where $\min(r_i) = 0.0$ and $\max(r_i) = 1.0$.

Fig. 10: Sliding window algorithm [18]

learning algorithms as shown in Table 5. Our results indicate that the accuracy of this technique is not that high as it was originally developed to preserve local details of a file ( [18])while the size of file affects localness a lot. In our case file sizes vary from around 0.5Kb to 53.7Mb which adversely affect the results. Despite worse results it is still easier to distinguish between benign executables and malware than between malware from different time slices. Also we can see that ANN is better in *Benign vs Malware_000* dataset, C4.5 in *Benign vs Malware_207* and *Malware_000 vs Malware_207* datasets.

Table 5: Classification accuracy based on features from bytes n-gram randomness profiles, in %

| Dataset | Naive Bayes | BayesNet | C4.5 | k-NN | SVM | ANN | NF |
|---|---|---|---|---|---|---|---|
| | | | **All features** | | | | |
| Bn vs Ml_000 | 69.9 | 60.4 | 76.9 | 75.6 | 78.3 | 78.3 | 74.8 |
| Bn vs Ml_207 | 70.3 | 68.2 | 75.8 | 75.6 | 72.1 | 71.6 | 68.2 |
| Ml_000 vs Ml_207 | 50.1 | 64.0 | 68.1 | 64.7 | 58.1 | 60.1 | 58.2 |

Also it should be noted that we did not use feature selection methods as in the case of PE32 header features. Both Information gain and Cfs are not efficient due to the similarity of features and equivalence in importance for classification process. For the first dataset the Information Gain was in the range of 0.0473-0.0672 while for the second dataset it was in the range of 0.0672-0.1304 and for the last it was 0.0499-0.0725. Moreover, Cfs produces best feature subset nearly equal to full set. Therefore, we decided to use all features as there is no subset that could possibly be better than original one.

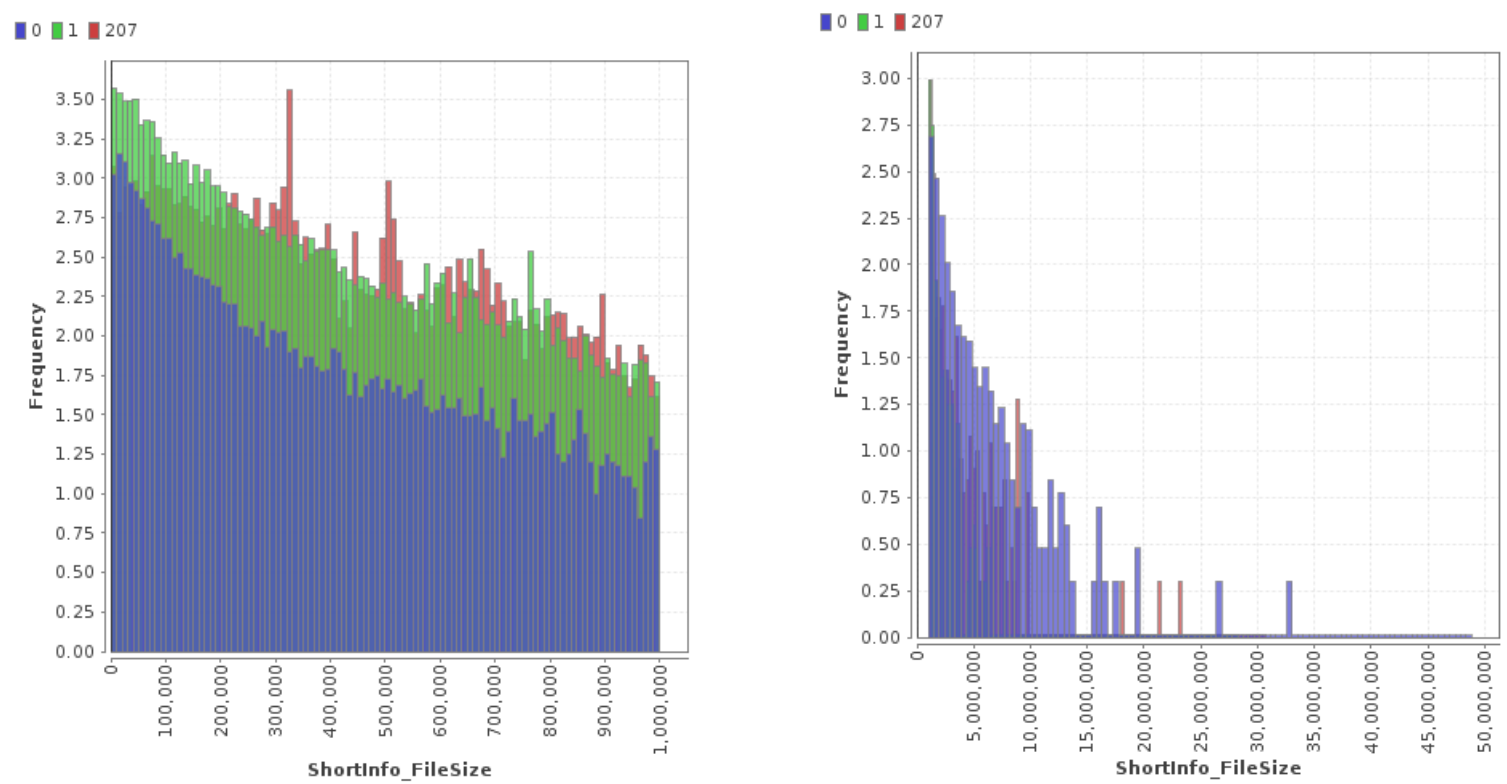

(a) Less than 1M (b) Bigger than 1M

Fig. 11: Distribution of file size values in Bytes for three classes

### 5.1.3 Opcode n-gram

Opcode n-gram consists of assembly instructions which construct the executable file. The main limitation of this method is that in order to gain opcodes we need disassemble an application which sometimes fails to give correct opcodes due to different anti-disassembly and packing techniques used in executables hence we filtered out this kind of files from our dataset. We extracted 100 most common 3- and 4-grams from each of three file sets in our dataset. Then we extracted a set of 200 most common n-grams - which are called feature n-grams - to build a presence vector where value 1 was assigned if a certain n-gram from feature n-grams is present in top 100 most used n-grams of the file. Table 6 represents results of feature selection performed on the dataset with 3-grams. As can be seen the first two pair of datasets have a lot of common n-grams, while selected n-grams for the third pair of dataset is totally different. For Information Gain the threshold of 0.1 was used for both benign and malware datasets, while for the last set we used InfoGain of 0.02.

These data were passed to machine learning algorithms and results are shown in Tables 7 and 9. As can be seen C4.5 performed well and had the highest accuracy almost in all experiments. Also feature selection significantly reduced the number of n-grams from 200 down to 10-15, while overall accuracy on all methods did not dropped significantly. In fact, Naive Bayes performed even better that can be justified by reduced complexity of the probabilistic model. Also NF showed much better accuracy in comparison to other methods when using all features to distinguish between two malware datasets which can be linked to non-linear correlation in the data that are circumscribed in the Gaussian fuzzy patches.

Further, we investigated if there is any correlation between n-grams in files that belong to both benign and malicious classes. We extracted relative frequency of each n-gram according to the following formula $h_{n-gram} = \frac{N^{Class}_{files \in n-gram}}{N^{Class}_{files}}$, where $N^{Class}_{files \in n-gram}$ indicates number of files in class that has n-gram and $N^{Class}_{files}$ is a total number of files in this class. The results for 3-gram is depicted in the Figure 12.

Table 6: Feature selection on 3-gram opcode features. Bold font denotes features that present in both datasets that include nenign samples

| Benign vs Malware_000 | | Benign vs Malware_207 | | Malware_000 vs Malware_207 | |
|---|---|---|---|---|---|
| **Information Gain** | | | | | |
| merit | attribute | merit | attribute | merit | attribute |
| 0.302483 | **int3movpush** | 0.298812 | **int3movpush** | 0.042229 | pushlcallpushl |
| 0.283229 | **int3int3mov** | 0.279371 | **int3int3mov** | 0.039779 | movtestjne |
| 0.266485 | **popretint3** | 0.227489 | **popretint3** | 0.037087 | callpushlcall |
| 0.236949 | **retint3int3** | 0.202162 | **retint3int3** | 0.031045 | pushpushlcall |
| 0.191866 | **jmpint3int3** | 0.193938 | **jmpint3int3** | | |
| 0.134709 | callmovtest | 0.108580 | retpushmov | | |
| 0.133258 | movtestje | | | | |
| 0.115976 | callmovpop | | | | |
| 0.114482 | testjemov | | | | |
| 0.101328 | poppopret | | | | |
| 0.100371 | movtestjne | | | | |
| **Cfs** | | | | | |
| attribute | | attribute | | attribute | |
| movtestje | | movmovadd | | pushpushlcall | |
| **callmovtest** | | retpushmov | | movtestjne | |
| callmovpop | | xormovmov | | movmovjmp | |
| retint3int3 | | **callmovtest** | | jecmpje | |
| **popretint3** | | **popretint3** | | cmpjepush | |
| **pushmovadd** | | **pushmovadd** | | pushleacall | |
| **int3int3mov** | | **int3int3mov** | | callpopret | |
| **callmovjmp** | | **callmovjmp** | | leaveretpush | |
| **jmpint3int3** | | **jmpint3int3** | | pushmovadd | |
| **int3movpush** | | **int3movpush** | | pushcalllea | |
| | | | | callpushlcall | |
| | | | | callmovlea | |
| | | | | pushlcallpushl | |
| | | | | movmovmovl | |
| | | | | calljmpmov | |

Table 7: Classification accuracy based on features from opcode 3-gram, in %

| Dataset | Naive Bayes | BayesNet | C4.5 | k-NN | SVM | ANN | NF |
|---|---|---|---|---|---|---|---|
| **All features** | | | | | | | |
| Bn vs Ml_000 | 83.51 | 83.52 | **95.53** | 93.82 | 94.43 | 94.51 | 95.28 |
| Bn vs Ml_207 | 84.52 | 84.52 | **93.93** | 91.84 | 92.32 | 92.44 | 93.20 |
| Mn_000 vs Ml_207 | 63.73 | 63.73 | 81.21 | 78.64 | 75.42 | 76.64 | **83.13** |
| **Information Gain** | | | | | | | |
| Bn vs Ml_000 | 86.74 | 86.94 | 90.41 | **90.45** | 89.98 | 90.26 | 84.45 |
| Bn vs Ml_207 | 86.22 | 86.22 | 86.22 | 86.22 | 87.46 | **87.48** | 83.36 |
| Mn_000 vs Ml_207 | 63.19 | 62.55 | 71.19 | **71.89** | 69.54 | 67.36 | 69.14 |
| **Cfs** | | | | | | | |
| Bn vs Ml_000 | 87.79 | 88.66 | 91.15 | **91.22** | 90.90 | 90.82 | 85.31 |
| Bn vs Ml_207 | 86.24 | 86.33 | **89.92** | 89.73 | 89.17 | 89.34 | 81.58 |
| Mn_000 vs Ml_207 | 86.24 | 86.33 | **89.92** | 89.73 | 89.17 | 89.34 | 69.25 |

As a reference we took top 20 most frequent n-grams from benign class and found frequency of the corresponding n-grams from both malware datasets. It can be seen that the frequency does not differ fundamentally, yet n-grams for both malicious classes tend to have very close numbers in comparison to benign files. Moreover, there is a clear dependency between both malicious classes. We also can notice that most of the features selected from two datasets that includes benign samples are same. This highlights reliability of the selected 4-grams and generalization of this method for malware detection.

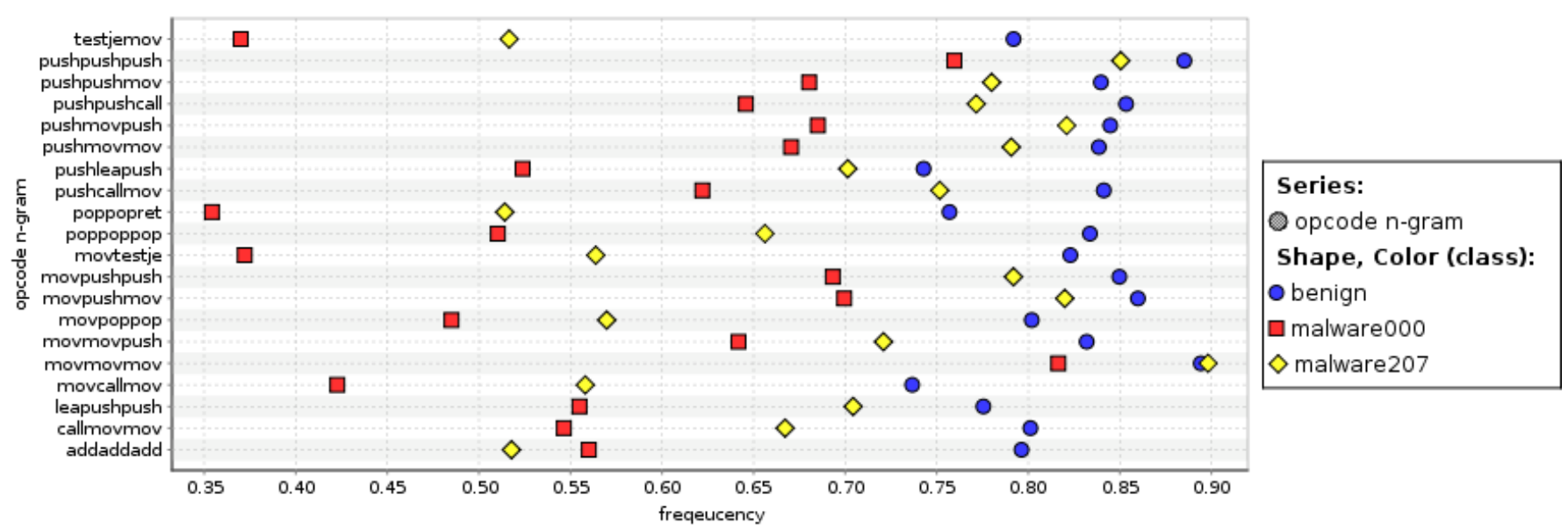

Fig. 12: Distribution of the frequencies of top 20 opcode 3-grams from benign set in comparison to both malicious datasets

Additionally, we studied 4-gram features and extracted 200 features as shown in Table 8. Similar to the 3-grams features selected in the Table 6 one can see that two first pairs of datasets have a lot of common features, while the last one provides a significantly different set. As in case with 3-grams we used Information Gain with threshold of 0.1 for both benign and the first malware dataset, while for the last malware set we used InfoGain of 0.02, which looks reasonable with respect to number of selected‘ features.

The classification performance is given in Figure 9. As can be seen, 3-grams can show a bit better result than 4-grams in case of distinguishing between benign and malware_000 or Benign and malware_207 with C4.5 classifier. At the same time 4-grams are better in order to distinguish between two malware datasets with C4.5 classifier. We can conclude that results are quite good, and can be used for malware detection. In our opinion results can be improved by extracting more features and usage of relative frequencies rather than pure vectors.

In contrary to 3-grams we can see that the histograms of 4-grams have fundamental differences when it comes to malicious and benign sets as it is depicted in Figure 13. We can see that the frequencies correspond to malware_000 and malware_207 datasets are nearly similar and are far from the frequencies detected for the benign class. Moreover, there is a clear and strong correlation between two malware datasets. So, we can conclude that in case of probabilistic-based models like Bayes Network and Naive Bayes the classification could be a bit better due to differences in likelihood of appearance, which can be also found in Tables 7 and 9.

Table 8: Feature selection on on 4-gram opcode features. Bold font denotes features that present in both datasets that include nenign samples

| Benign vs Malware_000 | | Benign vs Malware_207 | | Malware_000 vs Malware_207 | |
|---|---|---|---|---|---|
| **Information Gain** | | | | | |
| merit | attribute | merit | attribute | merit | attribute |
| 0.303209 | **int3int3movpush** | 0.295427 | **int3int3movpush** | 0.047452 | pushlcallpushlcall |
| 0.295280 | **int3movpushmov** | 0.286378 | **int3movpushmov** | 0.045860 | movpoppopret |
| 0.285608 | **int3int3int3mov** | 0.266966 | **int3int3int3mov** | 0.044750 | jepushcallpop |
| 0.258733 | **popretint3int3** | 0.229431 | **jmpint3int3int3** | 0.044573 | callpushlcallpushl |
| 0.241215 | **poppopretint3** | 0.224318 | **poppopretint3** | 0.038822 | cmpjepushcall |
| 0.233205 | **jmpint3int3int3** | 0.210289 | **popretint3int3** | 0.035731 | pushcallpopret |
| 0.220679 | **retint3int3int3** | 0.170367 | **retint3int3int3** | 0.030460 | pushcallpopmov |
| 0.185178 | **movpopretint3** | 0.148442 | **movpopretint3** | 0.028564 | movcmpjepush |
| 0.151337 | **movpushmovsub** | 0.116760 | **movpushmovsub** | 0.025813 | cmpjecmpje |
| 0.125703 | pushcallmovtest | 0.103841 | **movpushmovpush** | 0.024372 | leaveretpushmov |
| 0.104993 | **movpushmovpush** | 0.102730 | **movpushmovmov** | 0.023374 | pushpushpushlcall |
| 0.104416 | **movpushmovmov** | | | 0.022312 | pushcallpoppop |
| | | | | 0.021929 | movtestjepush |
| | | | | 0.020003 | pushpushleapush |
| **Cfs** | | | | | |
| attribute | | attribute | | attribute | |
| incaddincadd | | addaddaddadd | | leaveretpushmov | |
| **movpushmovsub** | | movmovpushpush | | callmovtestje | |
| jmpmovmovmov | | **movpushmovsub** | | jepushcallpop | |
| **pushcallmovtest** | | **pushcallmovtest** | | pushlcallpushlcall | |
| **int3int3int3mov** | | **int3int3int3mov** | | pushpushpushlea | |
| movpoppopret | | movxormovmov | | jecmpjecmp | |
| **jmpint3int3int3** | | pushlcallpushlcall | | movpoppopret | |
| movpopretint3 | | **jmpint3int3int3** | | pushcallmovpush | |
| **int3int3movpush** | | movpopretint3 | | pushmovmovcall | |
| **int3movpushmov** | | **int3int3movpush** | | movpopretint3 | |
| **poppopretint3** | | **int3movpushmov** | | cmpjepushcall | |
| addpushpushpush | | **poppopretint3** | | movleamovmov | |
| pushpushcalllea | | | | movmovjmpmov | |
| | | | | pushpushcalllea | |
| | | | | retnopnopnop | |
| | | | | movaddpushpush | |
| | | | | subpushpushpush | |

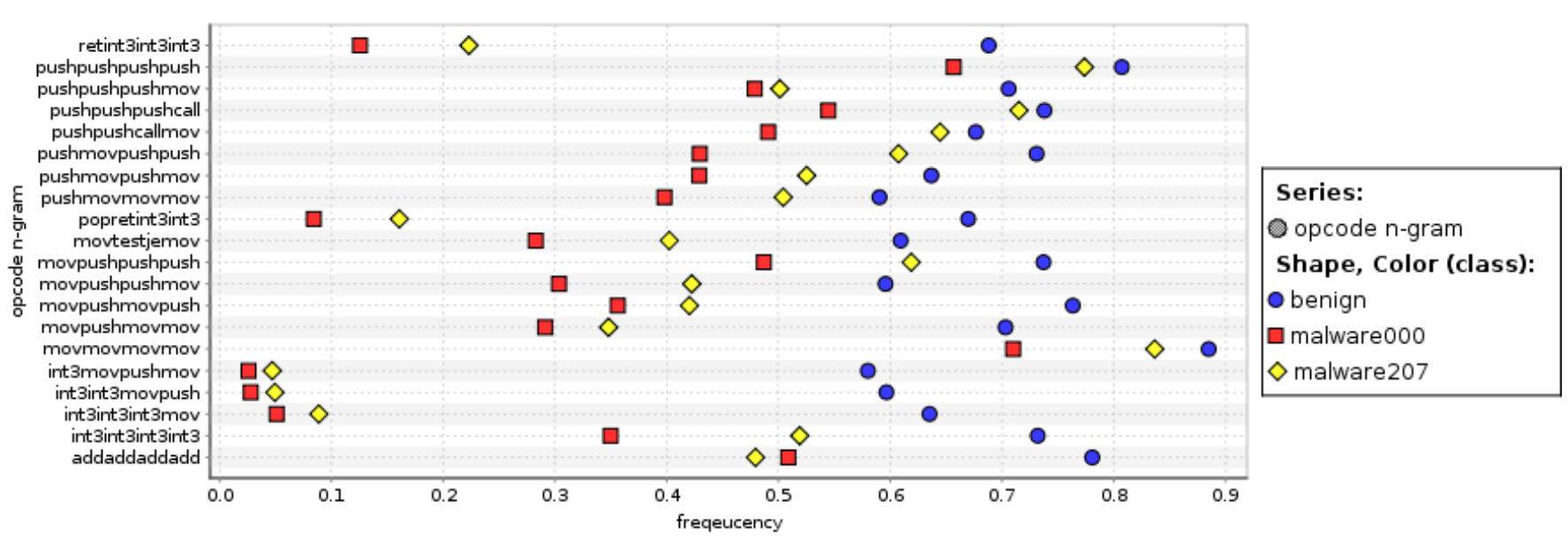

Fig. 13: Distribution of the frequencies of top 20 opcode 4-grams from benign set in comparison to both malicious datasets

Table 9: Classification accuracy based on features from opcode 4-gram, in %

| Dataset | Naive Bayes | BayesNet | C4.5 | k-NN | SVM | ANN | NF |
|---|---|---|---|---|---|---|---|
| **All features** | | | | | | | |
| Bn vs Ml_000 | 86.92 | 86.92 | 95.31 | 93.73 | 94.28 | 94.23 | **95.54** |
| Bn vs Ml_207 | 86.84 | 86.84 | 93.33 | 91.71 | 92.03 | 92.04 | **93.75** |
| Ml_000 vs Ml_207 | 64.90 | 64.90 | **81.58** | 78.98 | 74.98 | 75.77 | 78.80 |
| **Information Gain** | | | | | | | |
| Bn vs Ml_000 | 87.79 | 87.89 | **91.48** | 91.45 | 91.31 | 90.84 | 85.74 |
| Bn vs Ml_207 | 84.64 | 84.57 | **87.84** | 87.83 | 87.25 | 87.70 | 48.67 |
| Mn_000 vs Ml_207 | 62.73 | 63.20 | 69.96 | **70.25** | 68.40 | 67.24 | 68.90 |
| **Cfs** | | | | | | | |
| Bn vs Ml_000 | 89.63 | 89.63 | 91.51 | **91.52** | 91.52 | 90.76 | 84.95 |
| Bn vs Ml_207 | 86.41 | 86.64 | 89.36 | **89.48** | 89.16 | 89.12 | 81.13 |
| Mn_000 vs Ml_207 | 66.28 | 66.17 | 72.00 | **72.27** | 68.96 | 69.17 | 69.32 |

### 5.1.4 API call n-grams

API calls n-grams is the combination of specific operations invoked by the process in order to use functionalities of an operation system. In this study we used *peframe* to extract API calls from PE32 files. The bigger the n-gram size is the lower accuracy is possible to gain. The reason for this is that single API calls and their n-grams are far fewer in comparison with for example opcode n-grams. After extraction of API calls, we combined them into 1- and 2-grams. For each task we selected 100 most frequent features in a particular class and combined them into 200-feature vectors. Tables 10 and 11 presents results of machine learning evaluation on API call n-grams data.

As we can see ANN, k-NN and C4.5 are the best classifiers similar to previous results. It is also more difficult to distinguish between files from *malware_000* and *malware_207*. We gained quite high accuracy, but it is still lower than in related studies. It could be explained by the size of datasets: other studies datasets usually consist of several hundreds or thousands of files while our dataset has more than 110,000 files. After analysing feature selection results we decided not to include them in the results section since most of the features are similar in terms of distinguishing between malware and goodware. It means that there is large number of unique API calls that can be found once or twice in a file in contrary to the byte or opcode n-gram

Table 10: Classification accuracy based on API call 1-gram features, %

| Dataset | Naive Bayes | BayesNet | C4.5 | k-NN | SVM | ANN | NF |
|---|---|---|---|---|---|---|---|
| **All features** | | | | | | | |
| Bn vs Ml_000 | 90.79 | 90.79 | 93.39 | 93.47 | **93.51** | 93.43 | 82.44 |
| Bn vs Ml_207 | 87.18 | 87.18 | 90.94 | 91.03 | **91.37** | 91.23 | 81.28 |
| Ml_000 vs Ml_207 | 66.19 | 66.2 | **78.44** | 77.09 | 73.33 | 72.77 | 73.55 |

Table 11: Classification accuracy based on API call 2-gram features, %

| Dataset | Naive Bayes | BayesNet | C4.5 | k-NN | SVM | ANN | NF |
|---|---|---|---|---|---|---|---|
| **All features** | | | | | | | |
| Bn vs Ml_000 | 86.54 | 86.55 | 90.88 | 91.53 | **91.96** | 91.85 | 75.24 |
| Bn vs Ml_207 | 81.94 | 81.91 | 87.84 | **88.82** | 88.31 | 87.81 | 83.61 |
| Ml_000 vs Ml_207 | 62.31 | 62.31 | **73.69** | 73.17 | 70.27 | 69.45 | 70.08 |

We also studied the difference between frequencies distributions of API calls. Figure 14 sketches extracted API 1-grams from three datasets. One can see that there is a significant spread between numbers of occurrences in benign class in contrary to both malicious datasets. On the other hand, results for both malware datasets are similar, which indicates statistical significance of extracted features. It is important ho highlight that the largest scatter are in frequencies for *memset()*, *malloc()* and *free()* API calls. On the other hand, malicious programs tend to use *GetProcAddress()* function more often for retrieving the address of any function from dynamic-link libraries in the system.

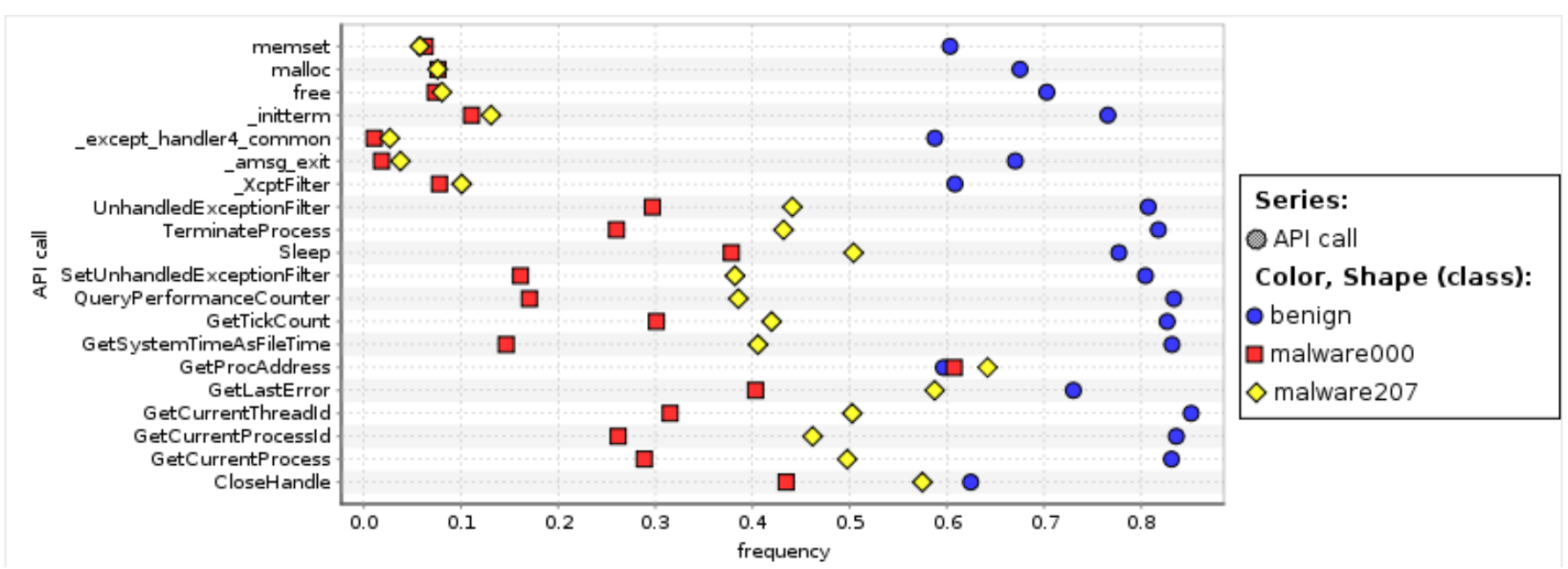

Fig. 14: frequencies of 20 most frequent API 1-grams for three different datasets

## 6 Conclusion

In this paper we presented a survey on applications of machine learning techniques for static analysis of PE32 Windows malware. First, we elaborated on different methods for extracting static characteristics of the executable files. Second, an overview of different machine learning methods utilized for classification of static characteristics of PE32 files was given. In addition, we offered a taxonomy of malware static features and corresponding ML methods. Finally, we provided a tutorial on how to apply different ML methods on benign and malware dataset for classification. We found that C4.5 and k-NN in most cases perform better than other methods, while SVM and ANN on some feature sets showed good performance. On the other hand Bayes Network and Naive Bayes have poor performance compared to other ML methods. This can be explained by negligibly low probabilities which present in a large number of features such as opcode and bytes n-grams. So, it can see that static-analysis using ML is a fast and reliable mechanism to classify malicious and benign samples considering different characteristic of PE32 executables. Machine Learning- aided static malware analysis can be used as part of Cyber Threat Intelligence (CTI) activities to automate detection of indications of compromise from static features of PE32 Windows files.